\documentclass[%
 reprint,
 amsmath,amssymb,
 aps,
]{revtex4-2}

\usepackage{graphicx}
\usepackage{dcolumn}
\usepackage{bm}
\usepackage[utf8]{inputenc}
\usepackage[T1]{fontenc}
\usepackage{mathptmx}
\usepackage{subfigure}

\usepackage{amsthm}

\newtheorem{assumption}{Assumption}

\begin{document}

\preprint{APS/123-QED}

\title{Charge management system based on disturbance observer sliding mode control for space inertial sensors}

\author{Fangchao Yang}
 \email{yangfc37@zhejianglab.com}
\affiliation{
Zhejiang Lab, Hangzhou 311121, People’s Republic of China.
}%
\author{Wei Hong}
 \email{hongwei@hust.edu.cn}
\affiliation{%
MOE Key Laboratory of Fundamental Physical Quantities Measurement, Hubei Key Laboratory of Gravitation and Quantum Physics, PGMF and School of Physics, Huazhong University of Science and Technology, Wuhan 430074, People's Republic of China.
}%
\author{Yujie Zhao}

\affiliation{%
College of Information Science and Engineering, Hunan Women's University, Hunan, 410004, People's Republic of China.
}%



\date{\today}

\begin{abstract}
Precision space inertial sensors are imperative for Earth geodesy missions, gravitational wave observations, and fundamental physics experiments in space. In these missions, free-falling test masses(TMs) are susceptible to parasitic electrostatic forces and torques, with significant contributions from the interaction between stray electric fields and TM charge. These effects can make up a sizable fraction of the noise budget. Thus, a charge management system(CMS) is essential in high-precise space-based missions. However, the operating environment for space charge control is full of uncertainties and disturbances. TM charge tracking precision is negatively affected by many physical parameters such as external charging rate, quantum yield, UV light power, etc. Those parameters are rarely measured and supposed to vary because of changes in solar activity, temperature, aging of electronic components and so on. The unpredictability and variability of these parameters affects the CMS performance in long-term space missions and must be evaluated or eliminated. This paper presents a simple physics-based model of the discharging process with high charging/discharging rate based on the geometry of inertial sensors. After that, a disturbance observer sliding mode control (DOSMC) is proposed for the CMS with parametric uncertainties and unknown disturbance to maintain the TM charge below a certain level and improve its robustness. The simulation results show that the DOSMC is able to force the system trajectory coincides with the sliding line, which depends neither on the parameters or disturbances. In this way, the DOSMC can effectively ignore the parameter perturbation and external disturbances. The control precision can reach 0.1 mV, which is superior to that of a classic proportional– integral–derivative controller and conventional sliding mode control.
\end{abstract}

\maketitle
\newcommand{\RNum}[1]{\uppercase\expandafter{\romannumeral #1\relax}}

\section{Introduction}
\begin{figure}
\includegraphics[width=250pt]{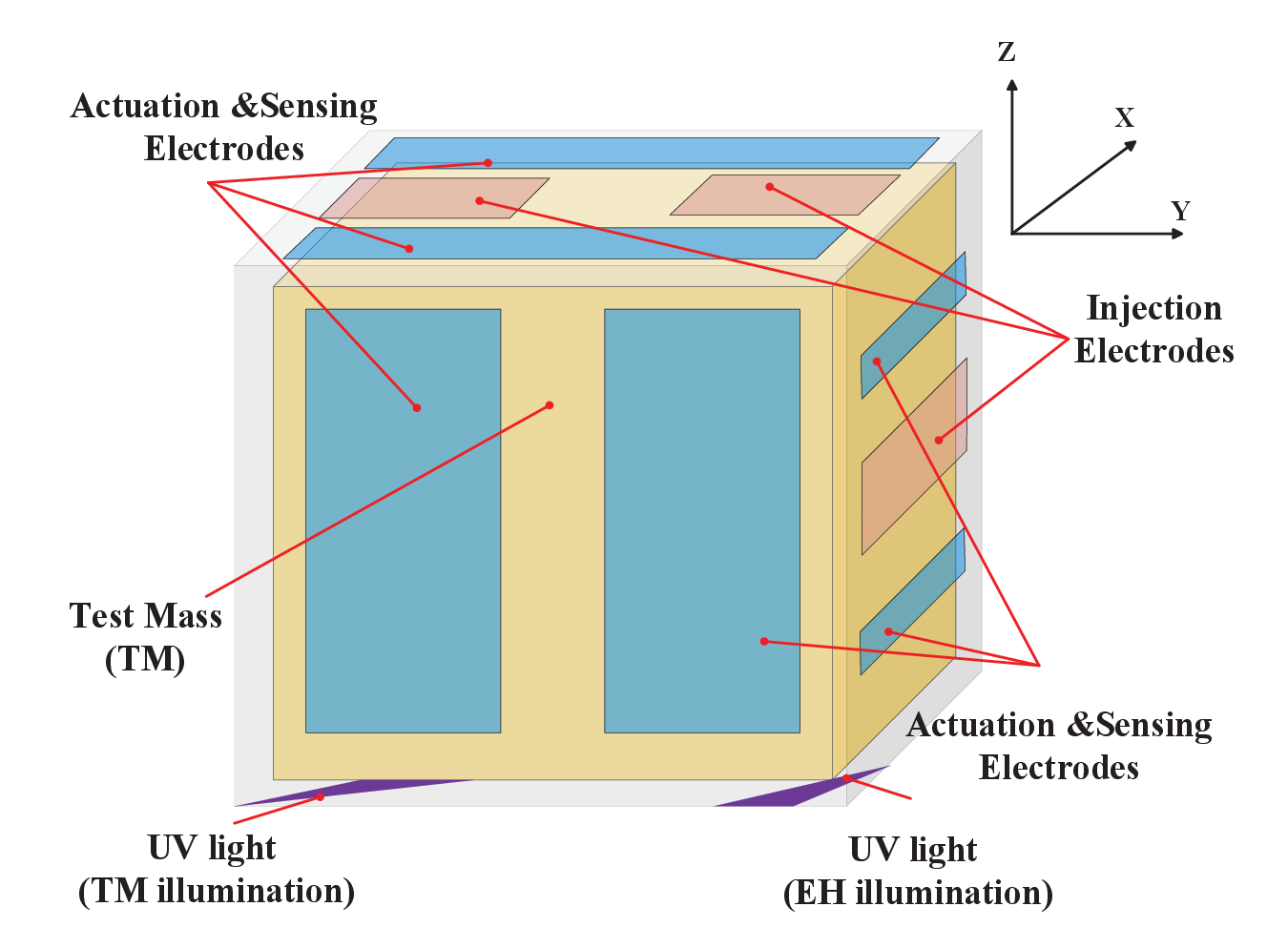}
\caption{\label{fig1} Schematic showing the geometry of an inertial sensor. The TM is indicated in yellow, the sensing and actuation electrodes are shown in blue, and the injection electrodes are shown in red. The UV light(purple) can be seen in the area between the -Z electrodes and TM.}
\end{figure}

Inertial sensors are an essential part of gravitational wave observatories\cite{LISA, TAIJI, TIANQIN, DECIGO2010}, satellite geodesy missions \cite{GOCE2010, Grace2019}, and several fundamental physics experiments\cite{STEP, Everitt2015, Touboul2022}. An inertial sensor(Figure~\ref{fig1}) includes an isolated free-floating test mass (TM) surrounded by six pairs of sensing and actuation electrodes and three additional pairs of electrodes responsible for injecting a high-frequency voltage on the TM for position and attitude readout. The completely electrically isolated TM inevitably accumulate an unwanted level of net charge and introduce additional acceleration noise as a result of galactic cosmic rays (GCRs), solar energetic particles (SEPs), and other unknown microscopic surface effects\cite{Jafry_1997,Wass_2005}. Thus, the TM charge must be kept close to a zero potential.

\subsection{\label{sec:level1}Charge Management System}
A charge management system(CMS) aims to control the TM charge in a contact-free manner by shining UV light and exploiting the photoelectric effect to avoid additional thermal noise introduced by mechanical contact. Electrons are liberated from gold-coated metallic surfaces using the photoelectric effect under illumination with UV light and move between the TM and its surroundings if the photon energy is greater than the energy required for an electron of an element to flee its valence band\cite{Olatunde2020}. The CMS is widely used in many space mission, including not only realized missions, such as GP-B\cite{GPB2015}, LISA Pathfinder(LPF)\cite{Armano_2017}, but also those future space mission, for example, LISA\cite{LISA}, TianQin\cite{TIANQIN}, and Taiji\cite{TAIJI}. Notably, in those realized missions, low-pressure mercury vapor lamps were used to generate UV light for discharge at a wavelength of 254 nm. Compared to mercury lamps, UV LEDs offer advantages in terms of size, electrical power efficiency, light weight, and low power consumption, and have been tested and identified as a new light source for future CMS\cite{Sun_2006, Hollington_2015, Yang2020, Pollack_2010}.

There are typically two types of charge control methods for high-precise space missions, namely fast discharge and continuous discharge\cite{Armano2018, Yang2023}. Both methods have been successfully used in the CMS of the LPF. Fast discharge is an open-loop control scheme, which is intended to quickly reduce the TM charge to the target level, disregarding the cause of any disturbance noise. During the mission, TMs are permitted to accumulate charge gradually over a few weeks until the absolute charge level becomes unacceptable. The TM is then rapidly discharged, typically within tens of minutes\cite{Armano2023}. Unlike fast discharge, continuous discharge is a closed-loop charge control method that aims to keep the TM charge consistently near zero by continuously illuminating the inertial sensor. The discharging rate is adjusted automatically by the controller based on TM net charge. In this way, the charging-discharging rates are balanced.

Compared to fast discharge, continuous discharge scheme is more convenient and precise as it is able to automatically adjust the UV light power. Additionally, the noise generated by the continuous discharge is lower, making it more suitable for future long-term space missions. However, there are still several complicated problems that will affect or disturb the continuous long-term TM charge control. First, although SEP events occur occasionally (several times per year), the expected TM charging rate caused by these events is approximately four to five times above the level generated by the flux of GCRs and can even reach as much as 70000 es$^{-1}$ at a peak flux\cite{Wass_2005}. Furthermore, the output power of the UV light generated by UV LEDs will decrease by as much as 66 ${\rm{\% }}$ for SET-240 devices and 10${\rm{\% }}$ for CIS-250 devices after each device is turned on and off for about 50 days\cite{Hollington_2017}. Finally, the properties of the gold-coated surfaces, such as quantum yield or work function, vary on different surfaces, or even at different times on the same surface, owing to changes in temperature, pressure, and air contamination\cite{Wass_2019}. Therefore, to solve the position tracking control problem and improve the control performance of CMS under the condition of parametric uncertainties for long-term space missions, the control system is critical.

\subsection{\label{sec:level2}Sliding mode control}
A proportional–integral–derivative (PID) controller is a control loop mechanism that employs feedback, is widely used in industrial control systems\cite{Bennett1993}, which is also used in continuous discharge in CMS. However, fixed-gain PID controllers have to be retuned for different situations of radiation environments, surface properties, or output power of UV light to ensure the control performance and robustness, which is almost impossible owing to parametric uncertainties in long-term space missions.

In\cite{Yang2023}, a model reference adaptive control (MRAC) is proposed for the CMS with parametric uncertainties to maintain the TM charge below a certain level, and the control precision can reach 0.1 mV under uncertainties. However, the settling time of MRAC under uncertainties is relatively long, leading to interruptions in space scientific experiments.

Sliding mode control(SMC) is a special kind of the variable structure control which has proven to be an effective robust control strategy for incompletely modeled or parametric uncertainties since its first appearance in 1950s\cite{Utkin_1977}. The essence of SMC is to force the system states to operate on a prescribed space surface called as sliding-mode surface in a desired manner. Its state trajectory coincides with the sliding line, which depends neither on the system parameters nor on the disturbance; rather, it depends only on the sliding coefficient. In such a case, the reference signal can be tracked. Importantly, this control is discontinuous as it switches between two control functions\cite{Vsabanovic2011}. It has been used extensively in control systems due to its superior performance characterized by simplicity, fast dynamic response, and insensitivity to parameter perturbation and external disturbances\cite{Young1999,Gao1993}.Therefore, SMC is able to improve the performance of a closed-loop CMS when the system is aging or there are modeling uncertainties that result from unforeseen changes in the surface properties of a TM or from external disturbances, such as rapid changes in the radiation environment. Consequently, an SMC-based charge control method has the potential to solve the parametric variation problem for long-term space missions.

In this paper, we present a physical-based theoretical model of the discharging process with high charging/discharging rate based on the geometry of an inertial sensor together with a SMC-based charge control algorithm to address the issue of parametric uncertainty in CMS for high-precise space missions. The paper is structured as follows: first, we present a discussion of the theoretical charge control model that describes the discharging processes inside the inertial sensors.. After that, the disturbance observer-based SMC(DOSMC) for CMS that uses Lyapunov’s stability theory is proposed. Subsequently, a closed-loop control system with parametric variations, including external charging rate, UV light power, and quantum yield, is designed, and the simulation results of the system performance are provided. Finally, the conclusions and discussion are presented.

The contributions of this study are as follows: (a) proposal of a simplified mathematical model of charge actuator suitable for SMC, and (b) design of a closed-loop control system based on DOSMC to verify the tracking performance and robustness for charge control.
\section{Modeling and Analysis}
\subsection{\label{sec:level3} Theoretical derivation}
In an inertial sensor, each TM surface and the opposing electrode in the electrode housing(EH) form a simple parallel plate capacitor. by radiating UV light onto the TM or EH surfaces, photoelectrons are released from these surfaces via the photoelectric effect. Then the successful extraction and transfer of an electron from one surface to another against an opposing voltage depends on the kinetic energy perpendicular to the plates. The number of photoelectrons that can migrate from surface $i$ to surface $j$ per unit time can be simply written as\cite{Yang2023, Inchauspe2020}
\begin{equation}
{N_{\rm{ij}}} =  M_{\rm{i}} \cdot \frac{{\int_{\Delta {V_{\rm{ij}}}}^{{V_{\max }}} {f(\Delta {V_{\rm{ij}}},\nu ,\varphi_{i} )d\Delta {V_{\rm{ij}}}} }}{{\int_0^{{V_{\max }}} {f(\Delta {V_{\rm{ij}}},\nu ,\varphi_i )d\Delta {V_{\rm{ij}}}} }},
\label{eq:1}
\end{equation}
\begin{equation}
M_{\rm{i}} = \frac{{{P_{\rm{UV}}}\int_{\frac{{e\Delta {V_{\rm{ij}}} + \varphi_{\rm{i}} }}{h}}^{ + \infty } {k(\nu )d\nu } }}{{h\int_0^{ + \infty } {\nu k(\nu )d\nu } }}{L_{\rm{i}}}({\alpha _{\rm{i}}},{R_{\rm{i}}})Q{Y_{\rm{i}}},
\label{eq:2}
\end{equation}
where $P_{\rm{UV}}$ is the UV power injected into the inertial sensor; $\nu$ is the optical frequency of a photon; $k(\nu )$ indicates the distribution of photon energies emitted by the UV LED, which can be approximated by a Gaussian function; $\varphi_
{\rm{i}}$ represents the work function of surface $i$; $\Delta {V_{\rm{ij}}}$ is the potential difference between surface $i$ and surface $j$; ${L_{\rm{i}}}({\alpha _{\rm{i}}},{R_{\rm{i}}})$ represents the fraction of the total UV power absorbed by the $i$th surface, which depends primarily on the reflectivity $R_{\rm{i}}$ and the incident angle $\alpha _{\rm{i}}$; $Q{Y_{\rm{i}}}$ is the quantum yield of surface $i$, defined as the number of photoelectrons emitted per absorbed photon; $M_{\rm{i}}$ indicate the total photoelectron flux generated by the UV illumination; and $f(\Delta {V_{\rm{ij}}},\nu,\varphi_{\rm{i}} )$ indicates the energy distribution of the released photoelectrons, which is well described by a triangular form with the maximum energy of a photoelectron $E_{\rm{max}}$ equal to the difference between the UV photon energy and the surface work function\cite{Armano2018,Wass_2019},
\begin{equation}
f(\Delta {V_{\rm{ij}}}) =
\begin{cases}
 0       &                             {\Delta {V_{\rm{ij}}} < 0}   \\
 \frac{{2\Delta {V_{\rm{ij}}}}}{{neV_{\max }^2}}      &                     {0 < \Delta {V_{\rm{ij}}} \le m{V_{\max }}} \\
 \frac{2}{{e{V_{\max }}(1 - m)}}(1 - \frac{{\Delta {V_{\rm{ij}}}}}{{{V_{\max }}}})  &  {m{V_{\max }} < \Delta {V_{\rm{ij}}} \le {V_{\max }}} \\
 0           &                         {\Delta {V_{\rm{ij}}} > {V_{\max }}} \\
 \end{cases},
\label{eq:3}
\end{equation}
where the parameter $m$ defines the fraction of $E_{\rm{max}}$ at which the triangular distribution peaks.

Substituting Equations ~(\ref{eq:3}) into ~(\ref{eq:1}) and setting the total area of the triangular energy distribution as 1, the photoelectron flow rate from surface $i$ to surface $j$ can be rewritten as follows,
\begin{equation}
{N_{\rm{ij}}} = {a_{\rm{ij}}}\Delta {V_{\rm{ij}}} + {b_{\rm{ij}}}\Delta {V_{\rm{ij}}}^2 + {c_{\rm{ij}}},
 \label{eq:4}
\end{equation}
where
\begin{align}
{a_{\rm{ij}}} &=
 \begin{cases}
 0              &                 \Delta {V_{\rm{ij}}} < m{V_{\max }}\;or\;  \Delta {V_{\rm{ij}}} > {V_{\max }} \\
  - \frac{{2e{M_{\rm{i}}}}}{{(1 - m)(h\nu  - \varphi_{\rm{i}} )}}  &    m{V_{\max }} < \Delta {V_{\rm{ij}}} < {V_{\max }} \\
 \end{cases},
 \label{eq:5}  \\
 {b_{\rm{ij}}} &=
 \begin{cases}
 0                 &              \Delta {V_{\rm{ij}}} < 0  \; or \; \Delta {V_{\rm{ij}}} > {V_{\max }} \\
  - \frac{{{M_{\rm{i}}}{e^2}}}{{m{{(h\nu  - {\varphi _i})}^2}}}     &       0 < \Delta {V_{\rm{ij}}} < m{V_{\max }} \\
 \frac{{{M_{\rm{i}}}{e^2}}}{{(1 - m){{(h\nu  - {\varphi _i})}^2}}}   &    m{V_{\max }} < \Delta {V_{\rm{ij}}} < {V_{\max }} \\
 \end{cases},
 \label{eq:6} \\
 {c_{\rm{ij}}} &=
 \begin{cases}
 {M_{i}}      &     \Delta {V_{\rm{ij}}} < m{V_{\max }} \\
 \frac{{{M_{\rm{i}}}}}{{1 - m}}   &     m{V_{\max }} < \Delta {V_{\rm{ij}}} < {V_{\max }} \\
 0        &     \Delta {V_{\rm{ij}}} > {V_{\max }}  \\
 \end{cases}.
  \label{eq:7}
\end{align}

In order to acquire the total rate of change of the TM potential under UV illumination, the inertial sensor is supposed to be split into pairs of adjacent surfaces on the TM and EH, and is treated as parallel plates with a uniform electric field between them. The total TM charging rate can then be obtained by calculating the net photoelectron flow between each surface pair while considering any instantaneous electrical potential difference,
\begin{equation}
 \begin{aligned}
 \frac{{d{V_{\rm{TM}}}}}{{dt}} &= \frac{e}{{{C_{\rm{T}}}}}\sum\limits_{i = 1}^n {({N_{\rm{tm - i}}} - {N_{\rm{i - tm}}}} ) \\\
 &= \frac{e}{{C_{\rm{T}}}}\sum\limits_{i = 1}^n {[\Delta a_{\rm{i}} \cdot \Delta {V_{\rm{tm - i}}} +\Delta b_{\rm{i}} \cdot \Delta {V_{\rm{tm - i}}}^2 + \Delta c_{\rm{i}}]}
 \end{aligned},
 \label{eq:8}
\end{equation}
where
\begin{equation}
\Delta a_{\rm{i}} = {a_{\rm{ji}}} + {a_{\rm{ij}}},
\label{eq:9}
\end{equation}
\begin{equation}
\Delta b_{\rm{i}} = {b_{\rm{ji}}} - {b_{\rm{ij}}},
\label{eq:10}
\end{equation}
\begin{equation}
\Delta c_{\rm{i}} = {c_{\rm{ji}}} - {c_{\rm{ij}}}.
\label{eq:11}
\end{equation}
Here, $n$ is the total number of split pairs of adjacent surfaces on the TM and EH; $C_{\rm{T}}$ is the total capacitance of the TM with respect to its surroundings; $N_{\rm{tm - i}}$ indicates the photoelectron flow rate moving from the TM surface to EH surface $i$; $N_{\rm{i - tm}}$ is the photoelectron flow rate from EH surface $i$ to the TM surface; $\Delta {V_{\rm{tm - i}}}$ is the potential difference between the TM surface and EH surface $i$; and $\Delta a_{\rm{i}}$, $\Delta b_{\rm{i}}$, $\Delta c_{\rm{i}}$ represent the relationship between the coefficients $a_{\rm{ji}}$, $b_{\rm{ji}}$, and $c_{\rm{ji}}$, respectively.

Considering the applied AC and DC voltages on the sensing and injection electrodes, equation.~(\ref{eq:8}) can be rewritten as,
\begin{equation}
 \begin{aligned}
 \frac{{d{V_{\rm{TM}}}}}{{dt}} &= \frac{e}{{{C_{\rm{T}}}}}\Big[\sum\limits_{i = 1}^n {\Delta {c_{\rm{i}}}}  + \sum\limits_{i = 1}^{n_{2}-1} {(\Delta {a_{\rm{i}}}} {V_{\rm{TM}}} + \Delta {b_{\rm{i}}}{V_{\rm{TM}}}^2) \\\
  &+ \sum\limits_{i = n_{2}}^{n_{1}-1} {\Big(\Delta {a_{\rm{i}}}} ({V_{\rm{TM}}} - {V_{\rm{inj}}}) + \Delta {b_{\rm{i}}}{({V_{\rm{TM}}} - {V_{\rm{inj}}})^2}\Big)\\\
   &+ \sum\limits_{i = n_{1}}^n {\Big(\Delta {a_{\rm{i}}}} ({V_{\rm{TM}}} - {V_{\rm{i}}}) + \Delta {b_{\rm{i}}}{({V_{\rm{TM}}} - {V_{\rm{i}}})^2}\Big)\Big]
  \end{aligned},
  \label{eq:12}
 \end{equation}
where $V_{\rm{i}}$ indicates the potential of the four Z-axis sensing electrodes; $V_{\rm{i}}=+V_{\rm{bias}}$ and $V_{\rm{i}}=-V_{\rm{bias}}$ voltages are applied to the electrodes of the negative and positive Z-axes, respectively; and $V_{\rm{inj}}$ is the injection voltage applied to  the six injection electrodes. The split pairs of adjacent surfaces, denoted as number $n_{1}=n - n_{\rm{b}}+1$ to $n$, represent the Z-axis sensing electrodes and their opposing TM surfaces. The number of surface pairs from $n_{2} =n - n_{\rm{i}}-n_{\rm{b}}+1 $ to $n_{1}$ indicates the injection electrodes and their opposing TM surfaces, and the remaining surface pairs are marked as number 1 to $n_{2}-1$. Furthermore, $n_{\rm{b}}$, $n_{\rm{i}}$ and $n_{\rm{r}}$ represent the number of DC bias electrodes, injection electrodes, and remaining electrodes, respectively.

Since the charge requirement for numerous space applications is usually tens of millivolts, the TM potential is quite small compared to the DC bias voltages and injection voltages while being controlled by the CMS in space. Thus, equation (1) can be simplified as ordinary first-order differential equations:
\begin{equation}
{\dot V_{{\rm{TM}}}}= {A_{\rm{TM}}}{V_{{\rm{TM}}}} + {A_{{\rm{bias}}}}{V_{{\rm{bias}}}} + {B_{{\rm{bias}}}}{V_{{\rm{bias}}}}^2 + C,
\label{eq:13}
\end{equation}
where
\begin{align}
{A_{\rm{TM}}} &= \frac{e}{{{C_{\rm{T}}}}}\sum\limits_{i = 1}^n {\Delta {a_{\rm{i}}}} \label{eq:14}\\
 {A_{{\rm{bias}}}} &= \frac{e}{{{C_{\rm{T}}}}}(\sum\limits_{i = {n_1}}^{{n_3} - 1} {\Delta {a_{\rm{i}}}}  - \sum\limits_{j = {n_3}}^n {\Delta {a_{\rm{j}}}} ) \label{eq:15}\\
 {B_{{\rm{bias}}}} &= \frac{e}{{{C_{\rm{T}}}}}\sum\limits_{i = {n_1}}^n {\Delta {b_{\rm{i}}}} \label{eq:16} \\
 {A_{{\rm{inj}}}}& =  - \frac{e}{{{C_{\rm{T}}}}}\sum\limits_{i = {n_2}}^{{n_1} - 1} {\Delta {a_{\rm{i}}}} \label{eq:17} \\
 {B_{{\rm{inj}}}} &= \frac{e}{{{C_{\rm{T}}}}}\sum\limits_{i = {n_2}}^{{n_1} - 1} {\Delta {b_{\rm{i}}}} \label{eq:18} \\
  C &= {A_{{\rm{inj}}}}{V_{{\rm{inj}}}} + {B_{{\rm{inj}}}}{V_{{\rm{inj}}}}^2 + \frac{e}{{{C_{\rm{T}}}}}\sum\limits_{i = 1}^n {\Delta {c_{\rm{i}}}}, \label{eq:19}
\end{align}

Equation.~(\ref{eq:13}) has the solution:
\begin{equation}
{V_{\rm{TM}}}(t) = ({V_0+{V_{\rm{eq}}}}){e^{{A_{\rm{TM}}}(t - {t_0})}}+{V_{\rm{eq}}},
\label{eq:20}
\end{equation}
where ${V_{\rm{eq}}}=- \frac{{{A_{{\rm{bias}}}}{V_{{\rm{bias}}}} + {B_{{\rm{bias}}}}{V_{{\rm{bias}}}}^2 + C}}{{A_{\rm{TM}}}}$ is the TM equilibrium potential after being illuminated by UV lights, $V_0$ indicates the initial TM potential before discharging, and $t_0$ is the discharging start time. The charging rate is calculated at the time when the TM potential crosses zero, namely, $t_{\rm{cross}} = \frac{{\ln [ - {V_{{\rm{eq}}}}/({V_0} + {V_{{\rm{eq}}}})]}}{{A_{\rm{TM}}}} + {t_0}$, then $\left. {\frac{{d{V_{{\rm{TM}}}}}}{{d{\rm{t}}}}} \right|\begin{array}{*{20}{c}}{}  \\{t = {t_{{\rm{cross}}}}}  \\ \end{array} =  - {A_{\rm{TM}}}{V_{{\rm{eq}}}}$ provides the charging rate when the TM potential crosses zero.

In order to linearize the model of discharge actuator, Equation(~\ref{eq:13}) can be written in terms of the deviation of $V_{\rm{TM}}$ and $V_{\rm{bias}}$ from the desired set points ${V_{{\rm{TM0}}}}$ and ${V_{{\rm{bias0}}}}$. The procedure is as follows: The goal of charge control is to keep the TM potential below the required charge level. Consequently, when the TM charge meets this requirement, both the equilibrium potential and charging rate are close to zero, then,
\begin{equation}
 {\dot V_{{\rm{TM}}}} = {A_{{\rm{TM}}}}{V_{{\rm{TM0}}}} + {A_{{\rm{bias}}}}{V_{{\rm{bias0}}}} + {B_{{\rm{bias}}}}{V_{{\rm{bias0}}}}^2 + C= 0.
\label{eq:21}
\end{equation}
The right-hand side of Equation(~\ref{eq:21}) can be expanded into a Taylor series about (${V_{{\rm{TM0}}}}$, ${V_{{\rm{bias0}}}}$). The high-order terms are neglected such that
\begin{equation}
\begin{aligned}
{\dot V_{{\rm{TM}}}} &\approx&
\frac{{\partial {\dot V_{{\rm{TM}}}}}}{{\partial {V_{{\rm{TM}}}}}}\left|    \begin{small}{\begin{array}{*{20}{c}}
   {}  \\
   \begin{array}{l}
 {V_{{\rm{TM}}}} = {V_{{\rm{TM0}}}} \\
 {V_{{\rm{bias}}}} = {V_{{\rm{bias0}}}} \\
 \end{array}  \\

\end{array}}    \end{small}\right.({V_{{\rm{TM}}}} - {V_{{\rm{TM0}}}}) \\
&+& \frac{{\partial {\dot V_{{\rm{TM}}}}}}{{\partial {V_{{\rm{bias}}}}}}\left|\begin{small} {\begin{array}{*{20}{c}}
   {}  \\
   \begin{array}{l}
 {V_{{\rm{TM}}}} = {V_{{\rm{TM0}}}} \\
 {V_{{\rm{bias}}}} = {V_{{\rm{bias0}}}} \\
 \end{array}  \\
\end{array}} \end{small} \right. ({V_{{\rm{bias}}}} - {V_{{\rm{bias0}}}})
\end{aligned}.
\label{eq:22}
\end{equation}

Additionally, since our focus is on the trajectories close to (${V_{{\rm{TM0}}}}$, ${V_{{\rm{bias0}}}}$), the parameters are set as follows:
\begin{align}
 {V_{\rm{p}}} &= {V_{{\rm{TM}}}} - {V_{{\rm{TM0}}}}\label{eq:23} \\
 {V_{\rm{bp}}} &= {V_{{\rm{bias}}}} - {V_{{\rm{bias0}}}}\label{eq:24}.
\end{align}
Equation(~\ref{eq:21}) can then be approximated by the following linear equation:
\begin{equation}
u = {\dot V_{\rm{p}}} = {A_{{\rm{TM}}}}{V_{\rm{p}}} + {k_{{\rm{bp}}}}{V_{{\rm{bp}}}}
\label{eq:25}
\end{equation}
where $k_{{\rm{bp}}}={A_{{\rm{bias}}}} + 2{B_{{\rm{bias}}}}{V_{{\rm{bias0}}}}$. Since the parameters ${V_{{\rm{TM0}}}}$ and ${V_{{\rm{bias0}}}}$ can be measured and calibrated for a certain CMS, $V_{\rm{p}}$ then indicates the TM potential, ${\dot V_{\rm{p}}}$ is the TM charging/discharging rate and ${V_{{\rm{bp}}}}$ represents the applied DC bias. The error between desired TM potential ${V_{{\rm{d}}}}$ and measured TM potential ${V_{\rm{p}}}$ while being controlled by the CMS in space can be expressed as,
\begin{equation}
e = {V_{\rm{d}}} - {V_{\rm{p}}}
\label{eq:26}
\end{equation}
since the desired potential is near zero, Equation(~\ref{eq:25}) can also be rewritten as,
\begin{equation}
u = {k_{\rm{p}}}e + {k_{{\rm{bp}}}}{V_{{\rm{bp}}}}
\label{eq:27}
\end{equation}
where ${k_{\rm{p}}} =  - {A_{{\rm{TM}}}} > 0$.

A discharge simulation based on Equation(~\ref{eq:12}) was performed to analyze the behavior of the inertial sensor's discharging process. The parameters used in this simulation model are as follows: the inertial sensor is modeled as a system comprising 24 parallel-plate capacitors, each consisting of a TM surface and an opposing EH surface. The $n_{\rm{b}}=$4, $n_{\rm{i}}=$6, and $n_{\rm{r}}=$14 represent the numbers of DC bias electrodes, injection electrodes, and remaining electrodes, respectively. The UV light entered the inertial sensor through the corners of the lower z-face, angled at approximately $\pm $15 degree relative to the z-plane. The  UV light source has a wavelength of 250 nm, corresponding to a photon energy of approximately 5.0 eV. The properties of each individual gold-coated surface, including quantum yield $QY_{\rm{i}}$ and work function $\varphi_{\rm{i}}$, were primarily obtained from the experimental results we reported previously\cite{Wass_2019} and the in-flight measurement in LPF\cite{Armano2018}. The quantum yield is $1.3 \times {10^{ - 5}}$ e per absorbed photon for the EH surface and $3.1 \times {10^{ - 5}}$ e per absorbed photon for the TM surface. All the surfaces in the inertial sensor have a work function $\phi = 4.2$ eV and a simple triangular energy distribution for emitted photoelectrons peaking at approximately 0.2$E_{\rm{max}}$. Finally, at 100 kHz, 5 V AC and DC voltages were applied to the injection and sensing electrodes.

The percentage of light that each surface absorbed for a given illumination (TM or EH illumination) ${L_{\rm{i}}}({\alpha _{\rm{i}}},{R_{\rm{i}}})$ was acquired from the Geant4 based ray-trace simulation, which were primarily obtained from the results we reported previously\cite{Yang2023}.  For LED 1, which primarily illuminated the TM, 53.58$\%$ of the total UV light injected was absorbed by the TM surfaces, whereas 25.15$\%$ was absorbed by the EH surfaces. For LED 2 aimed at the EH, 17.25$\%$ of the total UV light injected was absorbed by the TM surfaces, whereas 60.83$\%$ was absorbed by the EH surfaces. Note that a significant fraction of the UV light was absorbed within the gaps between the housing and the electrodes. Most of the photoelectrons generated from these areas have no adjacent regions on the TM and are unlikely to influence the discharge. Therefore, they were ignored. Further, the corresponding structures, such as TM central recess and caging fingers, were ignored as well.

\begin{figure}
\includegraphics[width=250pt]{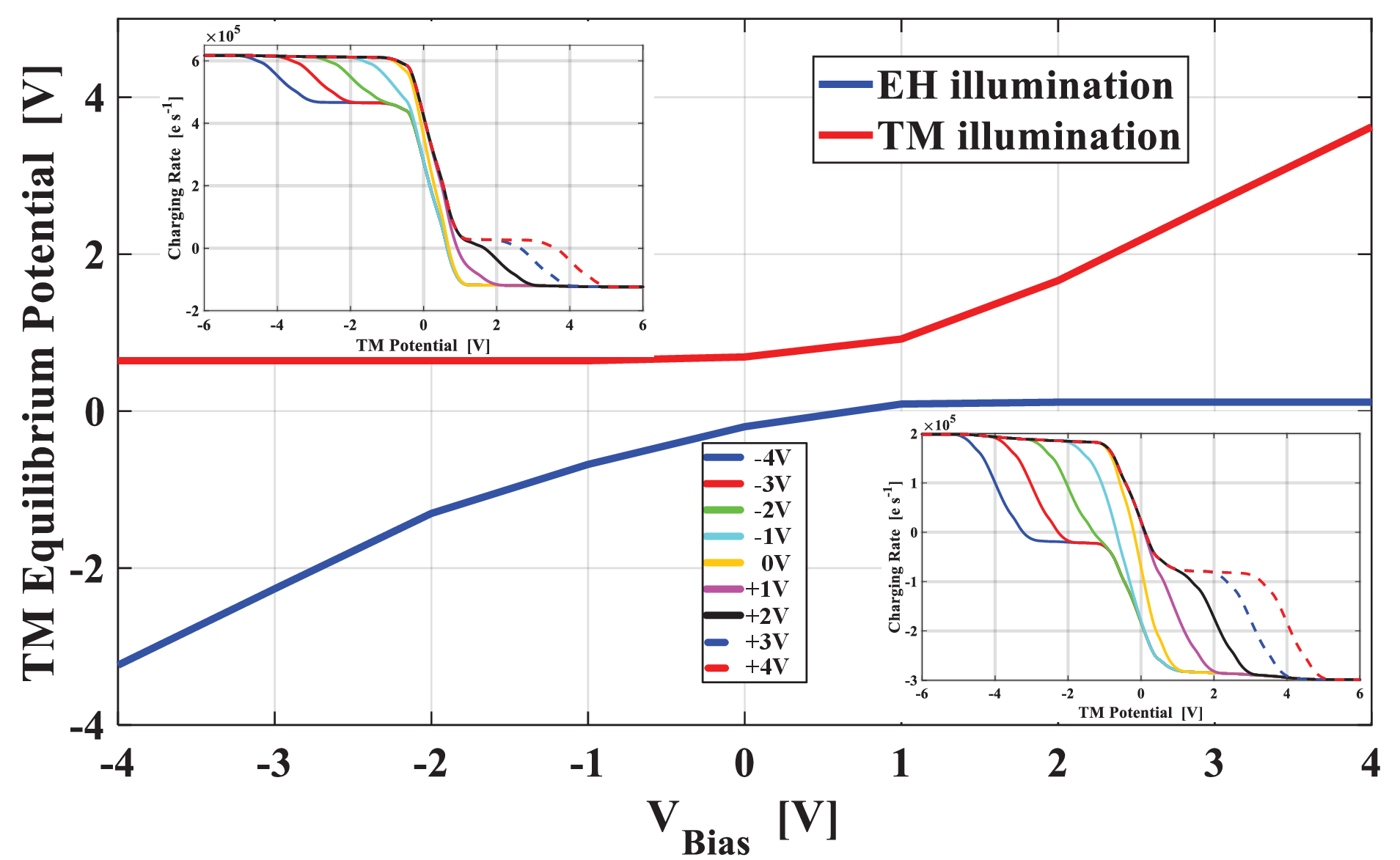}
\caption{\label{fig2} Model prediction of the discharge curves and equilibrium potentials for two different illumination cases. The DC bias gradually increases from -4.0 V to +4.0 V in a step of 1.0 V. Top-left: TM illumination. bottom-right: EH illumination.}
\end{figure}

Figure.~\ref{fig2} shows the discharge curves for both the TM and EH illuminations and equilibrium potentials for these two different illumination cases. These discharge curves represent the charging rate behavior expressed in terms of the charging rate as a function of the TM potential for positive and negative values of DC bias voltages. Equilibrium potentials are extracted from the zero-crossing of those discharge curves. For a DC bias voltage between -1.0 V and +1.0 V, two constant charging rate are observed in each illumination. At large negative potentials, all the photoelectrons originating from the TM are able to flow away from it, whereas none from the housing has sufficient energy to overcome the potential difference and reach the TM. The converse is also true at large positive potential differences. Moreover, there is a transition region near zero potential between the two constant levels. The point at which the curve crosses the apparent yield axis (the two-opposing-photocurrent balance and, therefore, zero apparent yield) is the equilibrium TM potential. As for other cases, namely a DC bias voltage between +1.0 V and +4.0 V and between -4.0 V and -1.0 V, one more transition region occurs when the TM potential approaches the corresponding DC bias voltage. In conclusion, by choosing an appropriate DC voltage, the polarity of the charging rate and equilibrium potential can be effectively shifted, and the desired direction of discharge and the target equilibrium potential can be obtained.

\subsection{\label{sec:level4} Model for SMC}

A high control gain is typically employed for SMC controllers to eliminate the adverse effects of disturbances and uncertainties and to ensure that the settling time of CMS is fast enough to enable continuous monitoring of scientific data. Therefore, the charging and discharging rates of charge actuator should be as high as possible when the TM potential approaches the desired potential value.

\begin{figure}
 \subfigure[Switching simultaneously the light illumination and the polarity of the bias voltage to achieve a maximum discharging rate]{
\includegraphics[width=250pt]{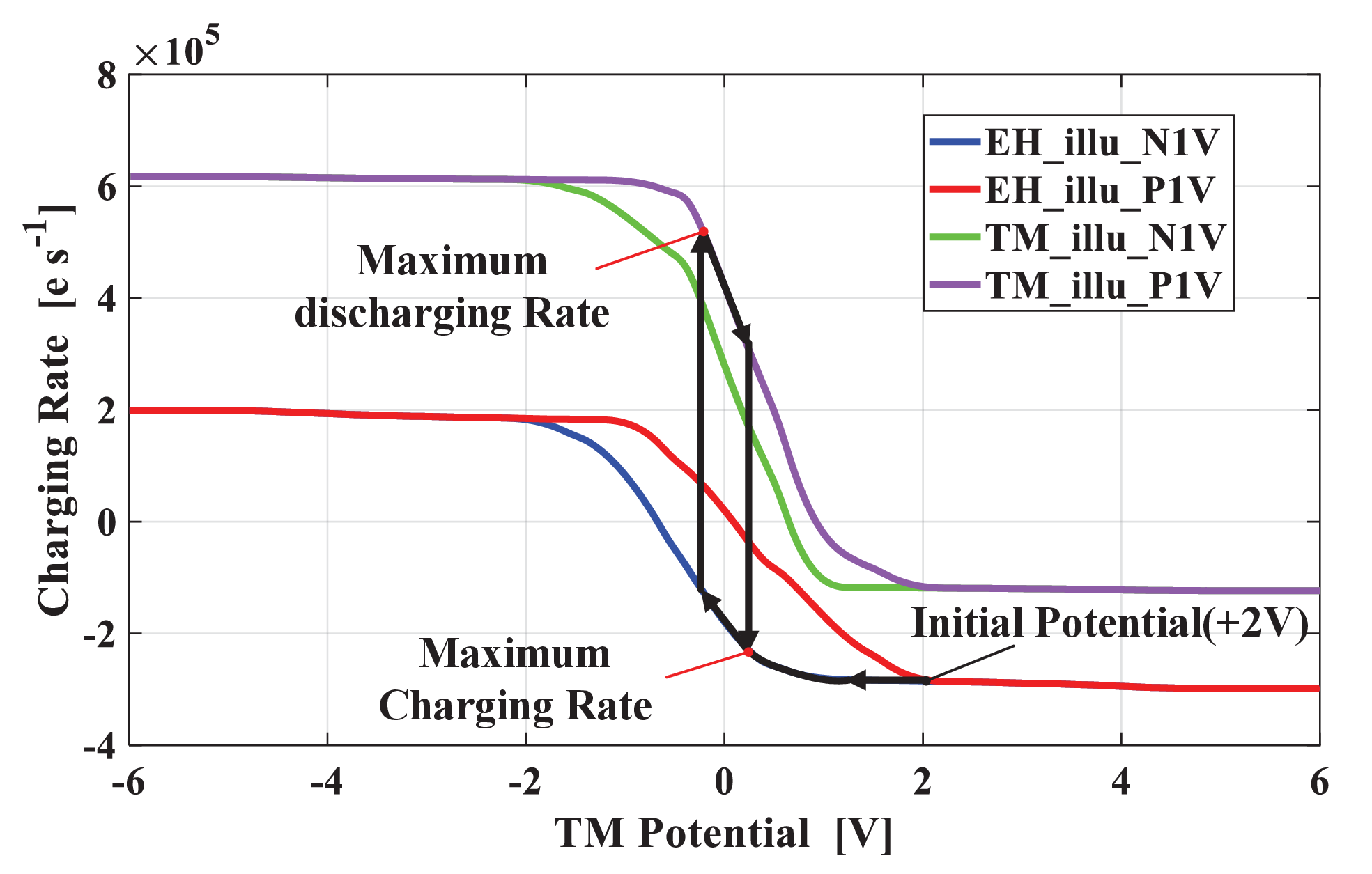}
}
\subfigure[Switching only the light illumination to achieve a high discharging rate]{
\includegraphics[width=250pt]{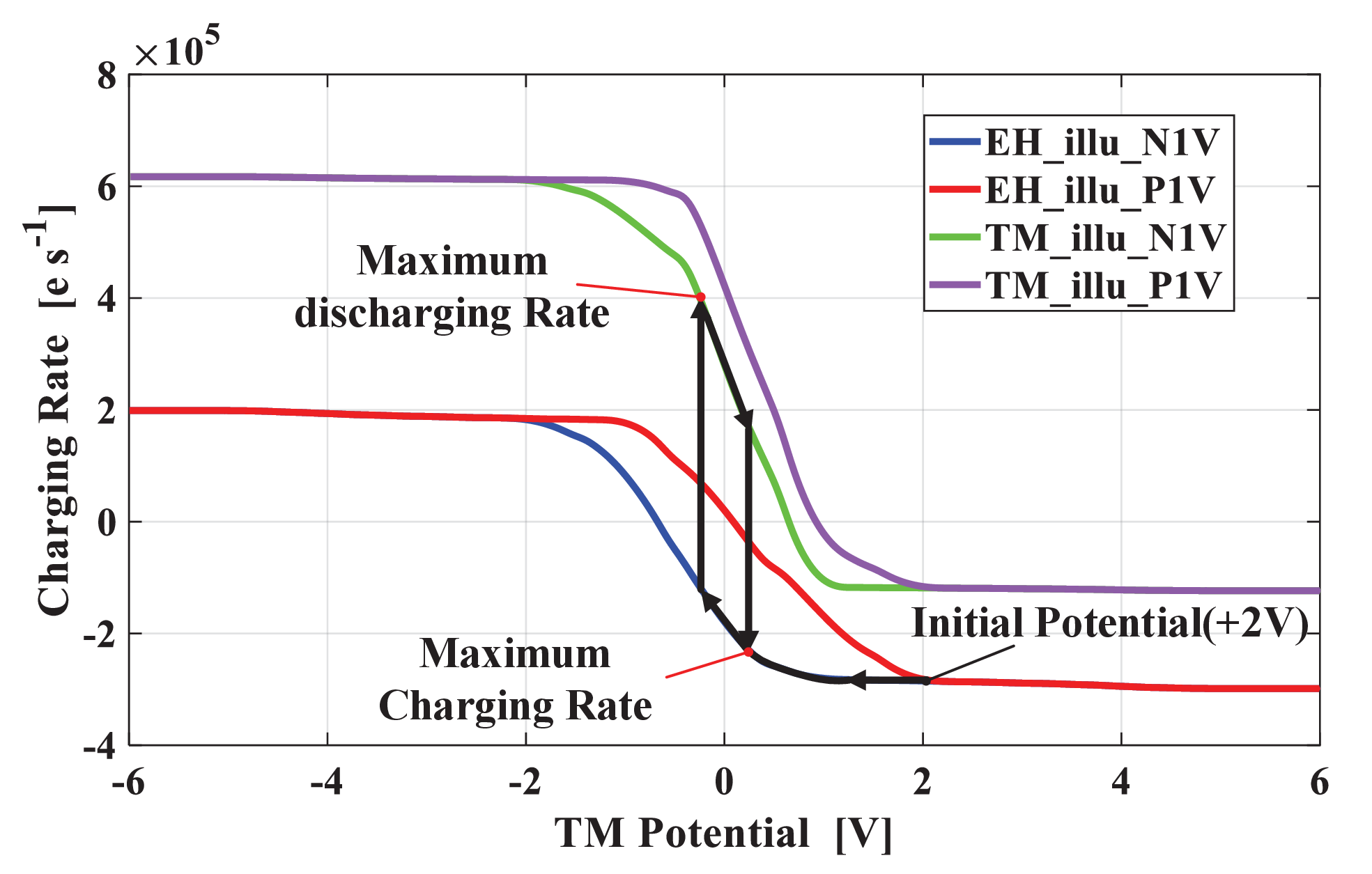}
}
\subfigure[Switching only the polarity of the bias voltage achieve a discharging rate near 0, which fails to meet the control requirement]{
\includegraphics[width=250pt]{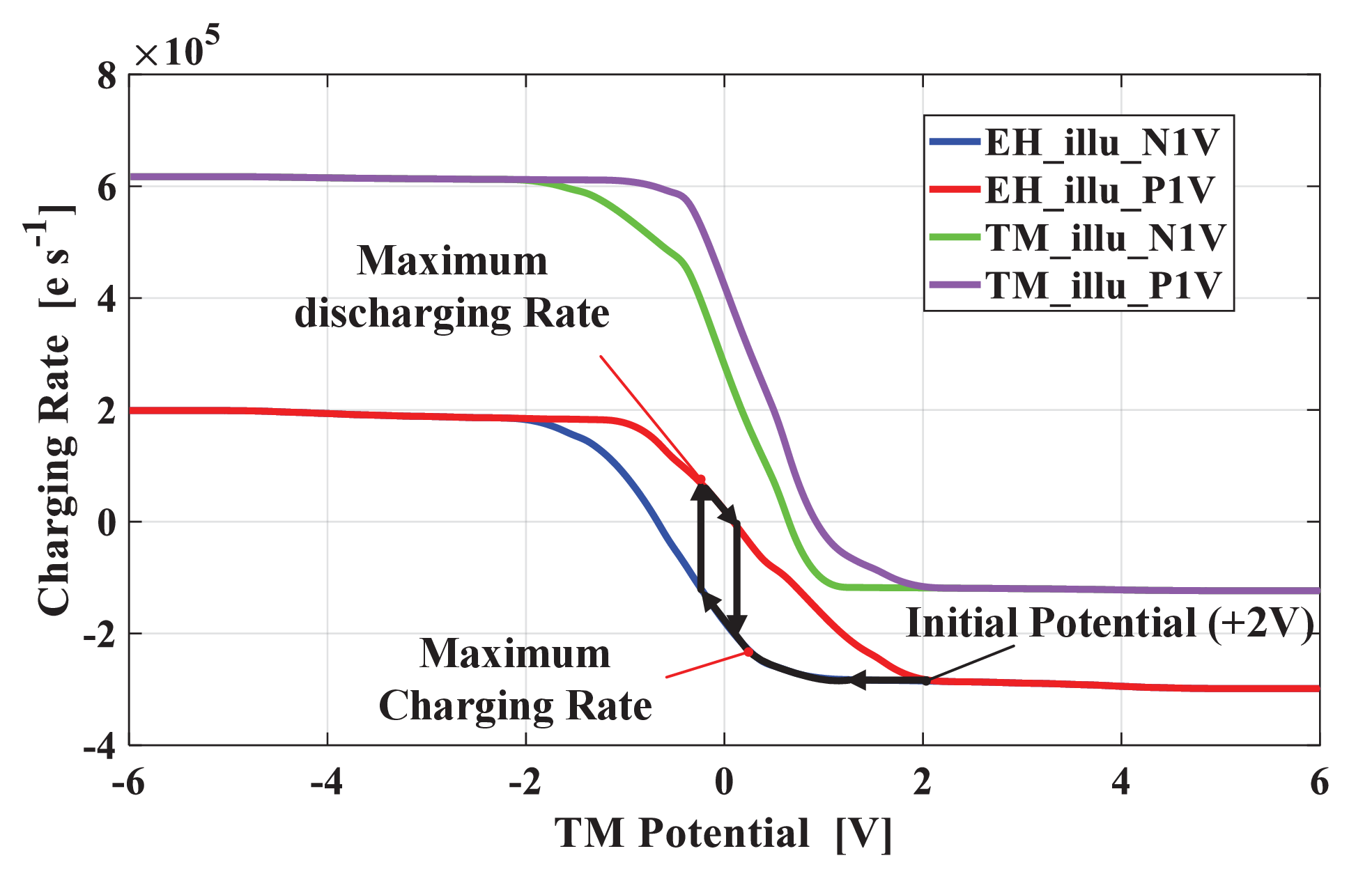}
}
\caption{\label{fig3} An example of the principle of switching charging and discharging rate when the initial potential of TM is positive(+2V). (a) Switch both the polarity of DC bias voltages and direction of UV light. (b)Switch only the direction of UV light. (c)Switch only the polarity of DC bias voltages. }
\end{figure}

\begin{figure}
 \subfigure[Switching simultaneously the light illumination and the polarity of the bias voltage to achieve a maximum charging rate]{
\includegraphics[width=250pt]{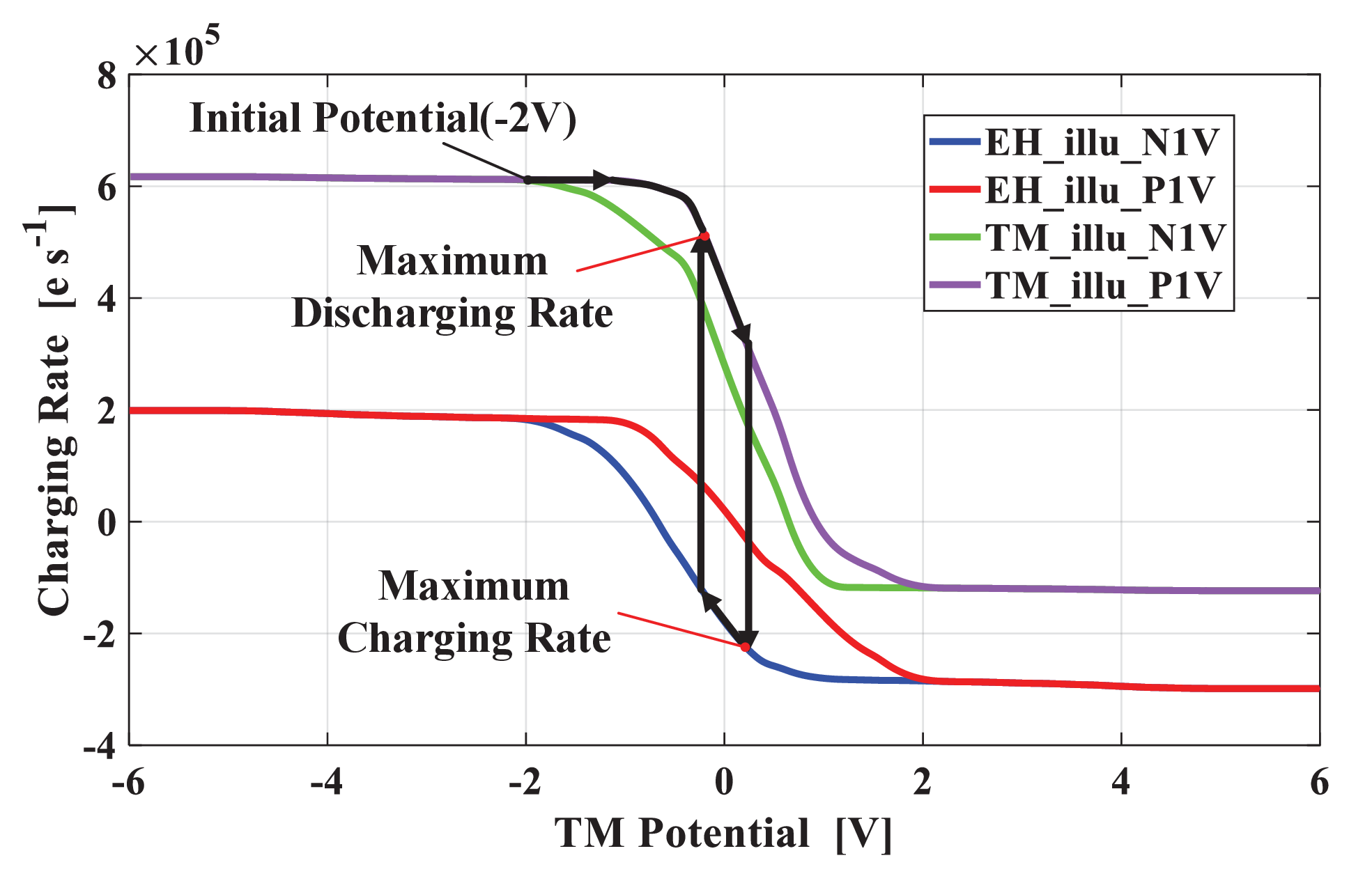}
}
\subfigure[Switching only the light illumination to achieve a high charging rate]{
\includegraphics[width=250pt]{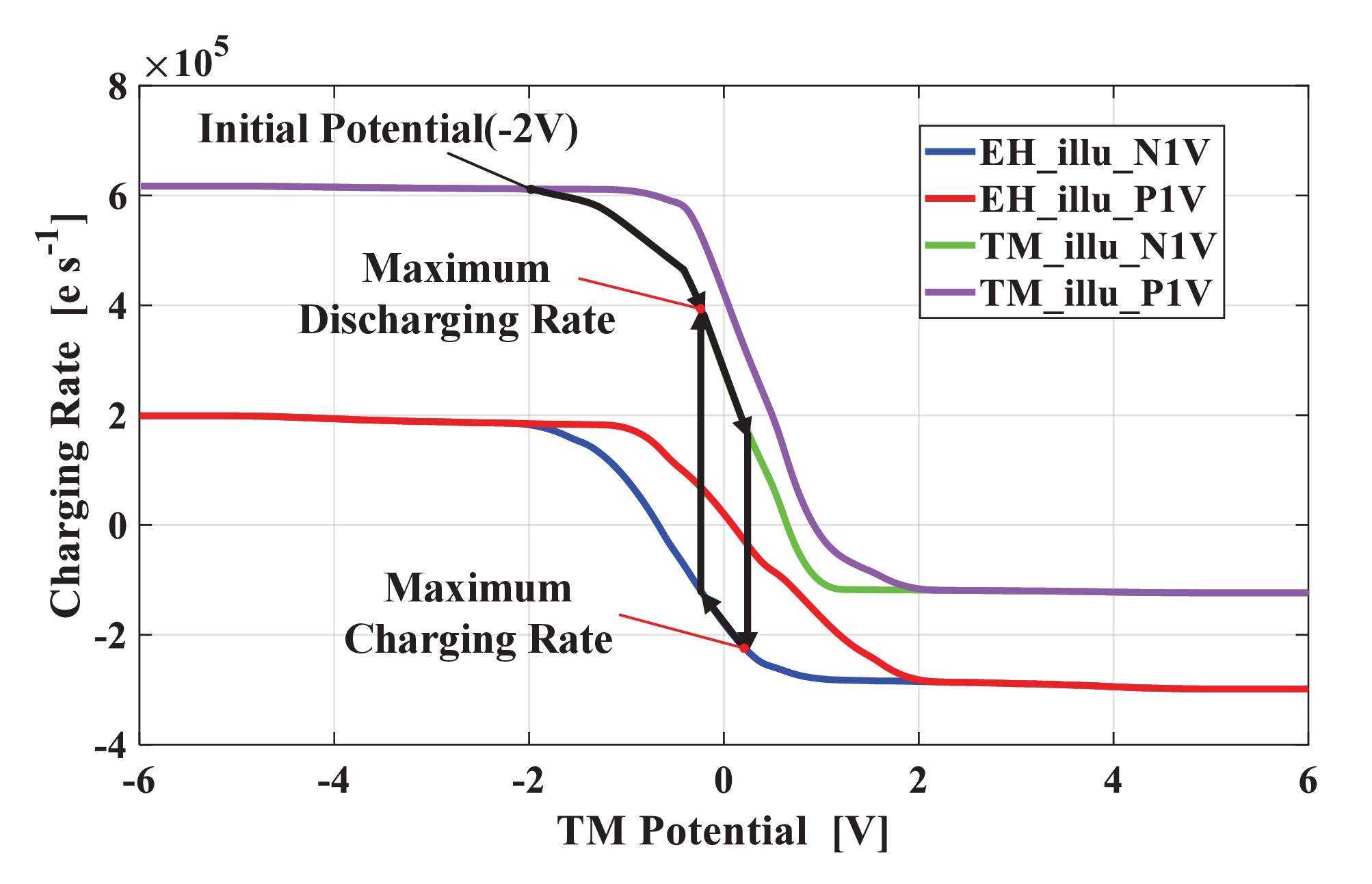}
}
\subfigure[Switching only the polarity of the bias voltage achieve a charging rate near 0, which fails to meet the control requirement]{
\includegraphics[width=250pt]{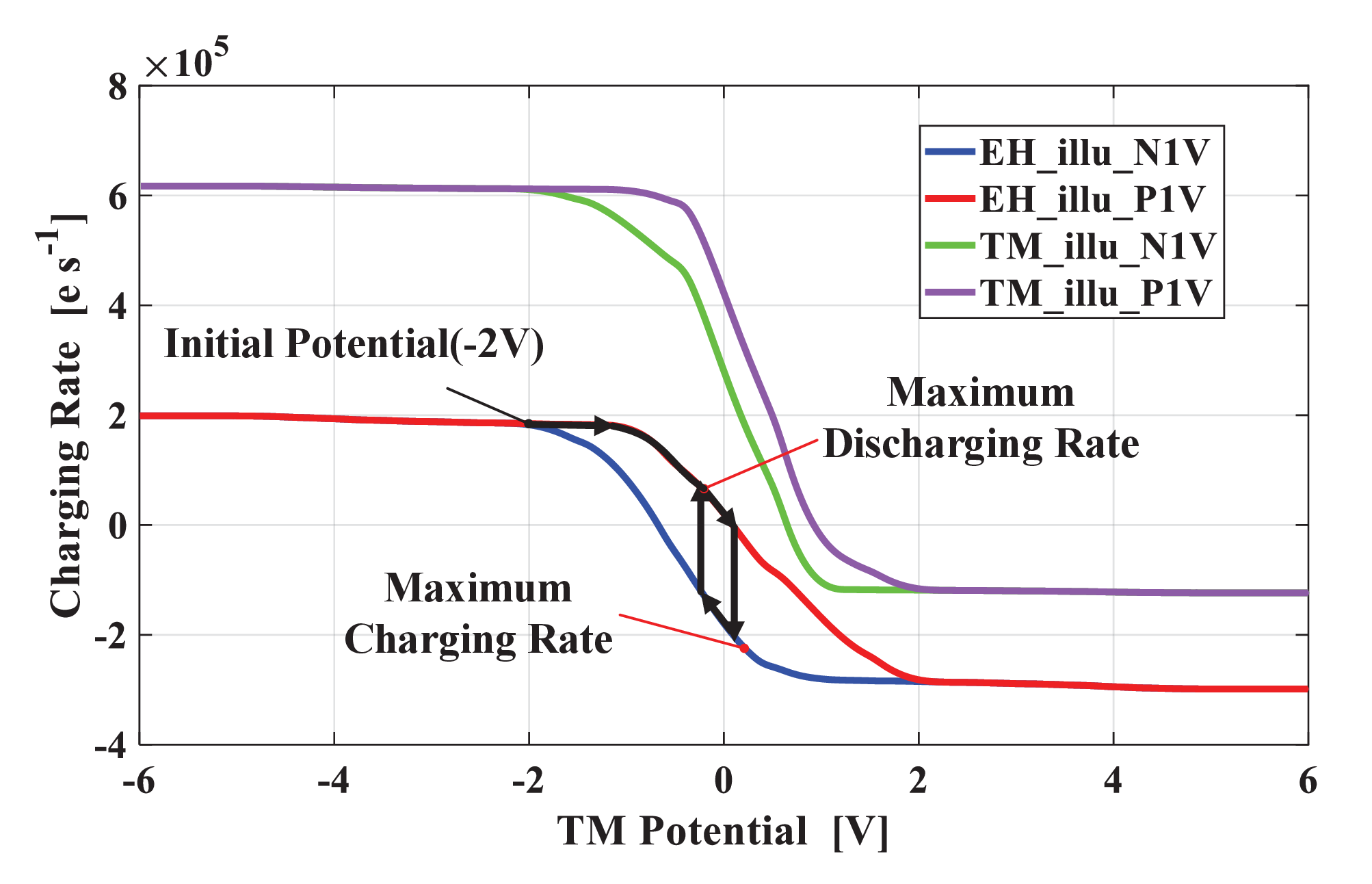}
}
\caption{\label{fig4} An example of the principle of switching charging and discharging rate when the initial potential of TM is negative(-2V). (a) Switch both the polarity of DC bias voltages and direction of UV light. (b)Switch only the direction of UV light. (c)Switch only the polarity of DC bias voltages. }
\end{figure}

Figure.~\ref{fig3} and Figure .~\ref{fig4} illustrate an example of the principle of maximize actuator output when the initial potential of TM is positive(+2V) and negative(-2V), respectively. The red, blue, green and purple curves represent the discharge curves for both the TM and EH illuminations with different DC bias obtained from the simulation results in Figure.~\ref{fig2}. The red and blue curves represent the discharge curves for EH illumination with applied DC biases of -1V and +1V, respectively. The green and purple curves indicate the discharge curves for TM illumination with DC biases of -1V and +1V applied.

In Figure~\ref{fig3}(a), TM potential gradually decreases along the blue curve to achieve maximum charging rate, $\sim$3$\times$ 10$^{5}$ e/s. Then after TM potential falls below the desired value(0V), the UV light illumination is redirected to TM and the bias voltage polarity is also switched from negative to positive to achieve maximum discharging rate(along purple curve), about 5$\times$ 10$^{5}$ e/s. This process is continuously repeated in charge control to ensure the rapid convergence of the TM charge. Figure~\ref{fig3}(b) and Figure~\ref{fig3}(c) convey essentially the same concept as Figure~\ref{fig3}(a) except that Figure~\ref{fig3}(b) only switches the illumination direction, and Figure~\ref{fig3}(c) only switches the polarity of the bias voltage. One can see that simultaneously switching the light illumination and the polarity of the bias voltage, or changing the illumination direction, can achieve a higher discharging rate. However, only switching the polarity of the bias voltage achieve a discharging rate near 0, which fails to meet the control requirement. Therefore, in the subsequent simulations, the charging/discharging rate from the former is used. It is worth noticing that since the equilibrium potential of all curves except the blue one is negative, for cases where the initial potential is positive, the potential can only change along the blue curve. This requires applying a negative bias voltage while illuminating EH. Figure~\ref{fig4} illustrates a comparable concept.

Based on the above discussion, to achieve the maximum charging and discharging rate, equation(~\ref{eq:25}) can be rewritten as,
\begin{equation}
u(t) = {\dot V_{\rm{p}}} =
\begin{cases}
 {k_{\rm{p}}}e -  \left|{{k_{{\rm{bp\_EH}}}}{V_{{\rm{bp\_N}}}}}\right|       &                             {{V_{\rm{d}}} - {V_{\rm{p}}} < 0}   \\
 {k_{\rm{p}}}e +  \left|{{k_{{\rm{bp\_TM}}}}{V_{{\rm{bp\_P}}}}}\right|       &                             {V_{\rm{d}}} - {V_{\rm{p}}} > 0\\
 \end{cases},
\label{eq:28}
\end{equation}
where ${k_{{\rm{bp\_EH}}}}$ and ${k_{{\rm{bp\_TM}}}}$  indicate the coefficients when the UV light is illuminating at EH and TM, respectively, ${V_ {{\rm{bp\_N}}}}$ and ${V_{{\rm{bp\_P}}}}$ represent the negative and positive DC bias voltages. Equation(~\ref{eq:28}) can be simplified as,
\begin{equation}
u(t) = {k_{\rm{p}}}e + \rho sign(e)
\label{eq:29}
\end{equation}
where
\begin{equation}
\rho =
\begin{cases}
 \left|{{k_{{\rm{bp\_EH}}}}{V_{{\rm{bp\_N}}}}}\right|       &                             {{V_{\rm{d}}} - {V_{\rm{p}}} < 0}   \\
 \left|{{k_{{\rm{bp\_TM}}}}{V_{{\rm{bp\_P}}}}}\right|       &                             {V_{\rm{d}}} - {V_{\rm{p}}} > 0\\
 \end{cases},
\label{eq:30}
\end{equation}

\section{Sliding mode control}
\subsection{\label{sec:level5} Conventional SMC for CMS}
SMC is a robust and effective control method for control systems suffering from parametric uncertainties and unpredictable interferences. It is developed from the idea of defining a switching function and a control law that drives the system states towards the sliding manifold, thereby reducing the tracking error to zero. Specifically, SMC essentially utilizes the discontinuous control signal to force the systems states to operate on a predesigned surface, i.e., the occurrence of the so-called sliding mode dynamic or sliding motion is enforced.

Due to the presence of various disturbances and uncertainties such as vibrations, solar energetic particles, and other unknown disturbances that deteriorate the charge control performance, a high control gain is typically employed for SMC controllers to eliminate the adverse effects of disturbances and uncertainties. However, switching gains that are too large will result in an undesired chattering phenomenon. To deal with such situations, the disturbance observer method has been proposed, which is able to effectively deal with the above problems by estimating the disturbances and performing compensation.

The total charging rate of a TM in space can be obtained by combining the charging rate caused by the radiation environment and the charging or discharging rate generated by the CMS\cite{Yang2023}:
\begin{equation}
{\dot V_{{\rm{p}}}}(t) = {A_{\rm{p}}}{V_{{\rm{p}}}}(t) + {B_{{\rm{p}}}}u(t)+ f,
\label{eq:31}
\end{equation}
where ${A_{\rm{p}}}$ is the actual decay rate of the charging rate, ${B_{{\rm{p}}}}$ is the attenuation coefficients of the charging rate generated by CMS. $f$ indicates the unknown external disturbance. Note that the system parameters ${A_{\rm{p}}}$ and ${B_{{\rm{p}}}}$ are likely to change as the circumstances change, for example, the radiation environment or UV light output power.

\begin{assumption}
The uncertain parameter $f$ in Equation(~\ref{eq:31}) is bounded and is given by $\left| f \right| < \left| {{f_{\max }}} \right|$ , where $\left| {{f_{\max }}} \right|$ is the maximum limit of the unknown disturbance.
\end{assumption}

Consider the sliding surface $s$ for first-order systems,
\begin{equation}
s = ce
\label{eq:32}
\end{equation}
where e is the tracking error signals for Equation(~\ref{eq:31}). $c > 0$ is arbitrary control variable.  In order to drive sliding surface to zero in finite time by using control input $u(t)$  for the convergence of tracking error and error dynamics to zero in the presence of bounded disturbance, the time derivative of surfacing can be expressed by substituting equation(~\ref{eq:31}), which yields
\begin{equation}
\dot s = c\left( {{{\dot V}_{\rm{d}}} - {A_{\rm{p}}}{V_{\rm{p}}}(t) - {B_{\rm{p}}}{k_{\rm{p}}}e - {B_{\rm{p}}}\rho sign(e) + f} \right)
\label{eq:33}
\end{equation}

In order to prove the stability of dynamics of Equation(~\ref{eq:33}) around $s=0$,  A positive definite Lyapunov function can be defined as,
\begin{equation}
V = \frac{1}{2}{s^2}
\label{eq:34}
\end{equation}
which must satisfy the conditions i.e. $V>0$ and $\dot V < 0$ for $s \ne 0$. Condition $V>0$ is obvious, to validate the second condition, the derivative of $V$ can be rewritten as,
\begin{equation}
 \begin{aligned}
 \dot V &= s\dot s =  - {B_{\rm{p}}}{k_{\rm{p}}}{c^2}{e^2} - {A_{\rm{p}}}{c^2}e{V_{\rm{p}}} - {B_{\rm{p}}}\rho {c^2}\left| e \right| - f{c^2}e \\
  &\le {c^2}[ - {B_{\rm{p}}}{k_{\rm{p}}}{e^2} + ({A_{\rm{p}}}{V_{\rm{p}}} - {B_{\rm{p}}}\rho  + \left| {{f_{\max }}} \right|)\left| e \right|] \\
 \end{aligned}
\label{eq:35}
\end{equation}
where $\left| {{f_{\max }}} \right|$ is the maximum limit of the unknown disturbance. ${A_{\rm{p}}}{V_{\rm{p}}}$ is the TM charging rate owing to the bombardment of the spacecraft by galactic cosmic rays and solar energetic particles. A Monte Carlo simulation of the interaction of GCR protons and He nuclei with the LISA spacecraft indicated TM charging rates of nearly 20 +e/s for solar maximum conditions, rising to 50 +e/s for solar minimum\cite{Wass_2005}. $\rho$ is the charging/discharging rate generated by UV light in CMS, typically 10$^{5}$ e/s\cite{Pollack_2010}, and attenuation coefficient  is 20${\rm{\% }}$ $\sim$ 40${\rm{\% }}$\cite{Hollington_2017}. Therefore, ${A_{\rm{p}}}{V_{\rm{p}}} <  < {B_{\rm{p}}}\rho $, Equation(~\ref{eq:35}) then can be rewritten as,
\begin{equation}
\dot V \le {c^2}[ - {B_{\rm{p}}}{k_{\rm{p}}}{e^2} + (\left| {{f_{\max }}} \right| - {B_{\rm{p}}}\rho )\left| e \right|]
\label{eq:36}
\end{equation}
by selecting the charging/discharging rate ${B_{\rm{p}}}\rho  > \left| {{f_{\max }}} \right|$, Equation(~\ref{eq:36}) becomes as,
\begin{equation}
\dot V \le  - {c^2}{B_{\rm{p}}}{k_{\rm{p}}}{e^2}
\label{eq:37}
\end{equation}

Substituting Equation(~\ref{eq:34}) in(~\ref{eq:37}), we have
\begin{equation}
\dot V + 2{c^2}{B_{\rm{p}}}{k_{\rm{p}}}V \le 0
\label{eq:38}
\end{equation}
Assuming a function $\alpha (t) > 0$, then Equation(~\ref{eq:38}) can be rewritten as
\begin{equation}
\dot V + 2{c^2}{B_{\rm{p}}}{k_{\rm{p}}}V =  - \alpha (t)
\label{eq:39}
\end{equation}
which has the solution:
\begin{equation}
 \begin{aligned}
 V(t) =& V(0)\exp ( - 2{c^2}{B_{\rm{p}}}{k_{\rm{p}}}t) \\
  -& \exp ( - 2{c^2}{B_{\rm{p}}}{k_{\rm{p}}}t)\int_0^t {\alpha (\tau )} \exp (2{c^2}{B_{\rm{p}}}{k_{\rm{p}}}t)d\tau  \\
 \end{aligned}
\label{eq:40}
\end{equation}
where $V(0)$ is the initial value of $V$, exp represents the exponential function to distinguish it from the error term e. It is obvious that the second term is greater than zero, then Equation(~\ref{eq:40}) becomes as,
\begin{equation}
V(t) \le V(0){\exp({ - 2{c^2}{B_{\rm{p}}}{k_{\rm{p}}}t}})
\label{eq:41}
\end{equation}
Substituting Equation(~\ref{eq:34}), we get
\begin{equation}
\left| {e(t)} \right| \le \left| {e(0)} \right|\exp ( - {c^2}{B_{\rm{p}}}{k_{\rm{p}}}t)
\label{eq:42}
\end{equation}
Therefore, we can prove that the tracking error $e(t)$ will reach the sliding surface in a finite time and tends to zero asymptotically as $t -  > \infty $. However, the CMS charging rate ${B_{\rm{p}}}\rho $ has to be high enough to achieve stability if the disturbance term $f$  becomes larger, which results in unwanted chattering phenomenon. To overcome such shortages, disturbance observer sliding mode control(DOSMC) based CMS is proposed in the next section.

\subsection{\label{sec:level6} DOSMC for CMS}
In this section a DOSMC control technique is proposed for CMS to mitigate the effect of unknown disturbance. It includes a conventional SMC and an internal state observer law that evaluating instantaneous value of the disturbance, the control parameters are then updated in control input.

Assuming observers of disturbance term $f$ and TM potential $V_{\rm{p}}$ can be expressed as follow
\begin{equation}
\begin{cases}
 \dot {\hat {f}} = {K_1}\left( {{V_{\rm{p}}}(t) - {{\hat V}_{\rm{p}}}(t)} \right)  \\
 {{\dot {\hat {V}}}_{\rm{p}}}(t) = {A_{\rm{p}}}{{\hat V}_{\rm{p}}}(t) + {B_{\rm{p}}}{u_{\rm{f}}}(t) + \hat f + {K_2}\left( {{V_{\rm{p}}}(t) - {{\hat V}_{\rm{p}}}(t)} \right)\\
 \end{cases}
\label{eq:43}
\end{equation}
where ${\hat {f}}$ and ${{\hat V}_{\rm{p}}}(t)$ are the estimations of $f$ and $V_{\rm{p}}(t)$, respectively; $K_1$ and $K_2$  are the tuning gains to be designed, ${u_{\rm{f}}}(t)$ indicates the control law to stabilize surface and surface dynamics, designed as follow
\begin{equation}
{u_f}(t) = u(t) - \frac{1}{{{B_p}}}\hat f
\label{eq:44}
\end{equation}
Substituting Equation(~\ref{eq:44}) in (~\ref{eq:31}) the surface dynamics becomes as
\begin{equation}
\dot s = c\left( {{{\dot V}_{\rm{d}}} - {A_{\rm{p}}}{V_{\rm{p}}}(t) - {B_{\rm{p}}}{k_{\rm{p}}}e - {B_{\rm{p}}}\rho sign(e) + f - \hat f} \right)
\label{eq:45}
\end{equation}
Therefore, a disturbance observer can be designed so that the estimator error $f - \hat f$ converges to origin asymptotically. In this case, it sharply reduces chattering and the required charging rate of CMS.

Similarly, to prove the stability of dynamics of Equation(~\ref{eq:45}). A positive definite Lyapunov function for DOSMC can be defined as $V = V_1 + V_2$, where
\begin{equation}
\begin{cases}
 {V_1} = \frac{1}{2}{s^2} \\
 {V_2} = \frac{1}{{2{K_1}}}{{\tilde f}^2} + \frac{1}{2}{{\tilde V}_p}^2 \\
 \end{cases}
\label{eq:46}
\end{equation}
$\tilde f$ and  ${{\tilde V}_p}$ are the estimator errors, namely
\begin{equation}
\begin{cases}
 \tilde f = f - \hat f \\
 {{\tilde V}_p} = {V_p} - {{\hat V}_p} \\
 \end{cases}
\label{eq:47}
\end{equation}

\begin{assumption}
The time derivative of disturbances is bounded and holds the condition ${\lim _{t \leftrightarrow \infty }}\dot f = 0$.
\end{assumption}

\begin{assumption}
The disturbance estimator error ${\tilde f}$ in Equation(~\ref{eq:47}) is also bounded, i.e., there exist an unknown positive constant $\left| {{{\tilde f}_{\max }}} \right|$ such that $\left| {\tilde f} \right| \le \left| {{{\tilde f}_{\max }}} \right|$.
\end{assumption}

Similarly, the derivative of $V_1$  can be written as:
\begin{equation}
 \begin{aligned}
 {{\dot V}_1} &= s\dot s = {c^2}( - {A_{\rm{p}}}{V_{\rm{p}}}e - {B_{\rm{p}}}{k_{\rm{p}}}{e^2} - {B_{\rm{p}}}\rho \left| e \right| - \tilde fe) \\
  &\le  - {B_{\rm{p}}}{k_{\rm{p}}}{c^2}{e^2} + {c^2}(\left| {{{\tilde f}_{\max }}} \right| - {B_{\rm{p}}}\rho )\left| e \right| \\
 \end{aligned}
\label{eq:48}
\end{equation}
Substituting Equation(~\ref{eq:43}) into (~\ref{eq:48}), we obtain
\begin{equation}
 \begin{aligned}
 {{\dot V}_2} =&  - \tilde f{{\tilde V}_p}(t) + {{\tilde V}_p}{A_{\rm{p}}}{{\tilde V}_p}(t) + {{\tilde V}_p}{B_{\rm{p}}}[u(t) - {u_f}(t)] \\
  +& {{\tilde V}_p}\left( {f - \hat f - \hat f - {K_2}{{\tilde V}_p}} \right) \\
  =& ({A_{\rm{p}}} - {K_2}){{\tilde V}_p}^2(t) \\
 \end{aligned}
\label{eq:49}
\end{equation}
The tracking error will reach zero in finite time and remains thereafter as long as the designed gain $K_1 >0$ ${K_2} > {A_{\rm{p}}}$ and the charging rate of actuator ${B_{\rm{p}}}\rho  > \left| {{{\tilde f}_{\max }}} \right|$ In addition, Since ${\tilde f}$ represents the error between $f$ and its estimated value $\hat f$, as long as $\hat f$ approaches $f$  through the implementation of a disturbance observer, the amplitude of ${\tilde f}$ can be very small, which is able to reduce the chattering.

\begin{figure}
\includegraphics[width=250pt]{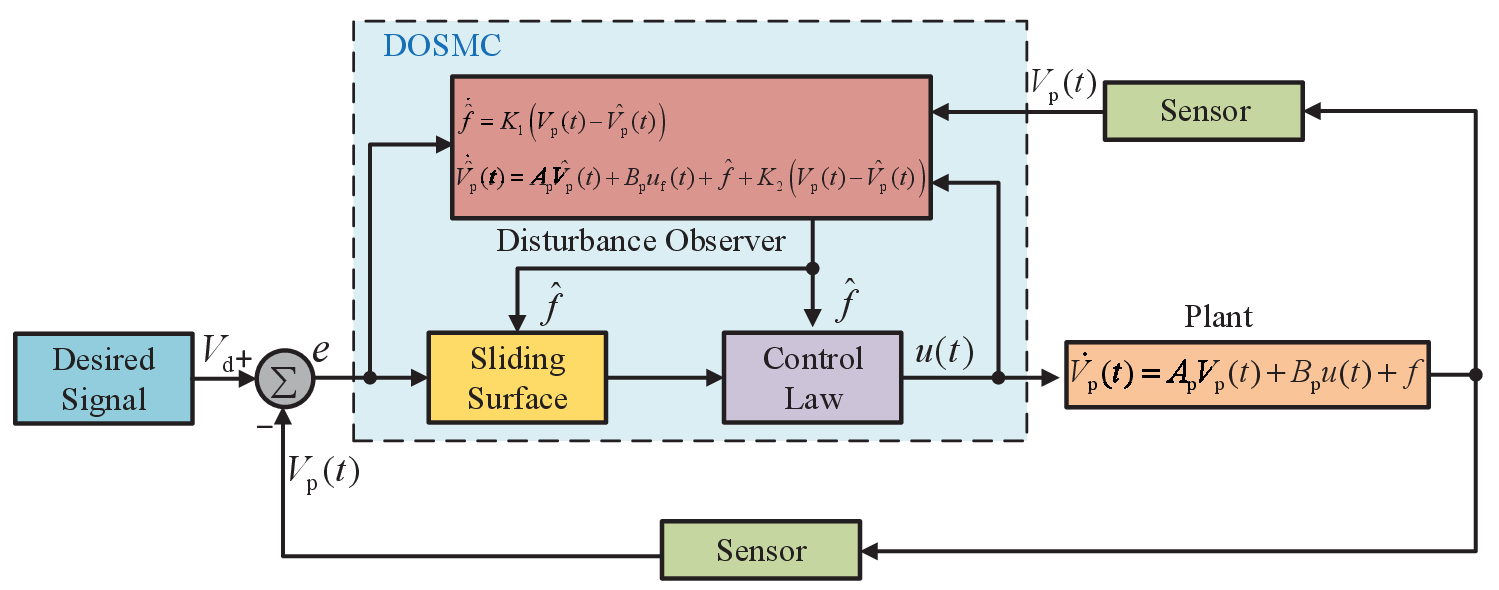}
\caption{\label{fig5} The block diagram of the proposed DOSMC-based CMS.}
\end{figure}

The block diagram of the proposed DOSMC-based CMS is shown in Figure~\ref{fig5}. The disturbance observer is an external loop for estimating the lumped disturbances from measurable variables and feeding the results back to the sliding-mode controller located in the inner loop in order to cause control actions to be taken to compensate for the influence of the disturbance.

\section{Results and Discussion}
\subsection{\label{sec:level7} Normal condition}
To verify the DOSMC performance, the performance of the CMS with stable parameters, which can be defined as normal conditions, should be evaluated. Under these conditions, it was assumed that the CMS parameters were known and remained stable. The responses of the TM potential under the designed DOSMC were then investigated and compared with those under the SMC and PID. The simulation process is described below. The parameters in the simulation are chosen as follows: The actual decay rate of the charging rate $A_{\rm{p}}$ was set to 3.13 $\times {10^{-7}}$ s$^{-1}$, which was equivalent to a charging rate of approximately +20 e/s. The UV attenuation coefficient $B_{\rm{p}}$ was set to 0.2. After several design iterations, we chose the estimation parameters to be $K_1 = 10000$ and $K_2 = 50$. Furthermore, assuming that the shot noise for the particle charging rates has a single-sided spectral density equal to +20 e/s/Hz$^{-1/2}$\cite{Araujo_2005}, the charge measurement noise is 1 fC/Hz$^{-1/2}$, and the electrode applied-voltage fluctuation is 30 $\mu$V/Hz$^{-1/2}$\cite{Armano_2017}.

\begin{figure}
 \subfigure[The closed-loop system response for SMC, DOSMC and PID. Top-left: magnified view of the rising edge of the control performance. Top-right: magnified view of the falling edge of the control performance.]{
\includegraphics[width=250pt]{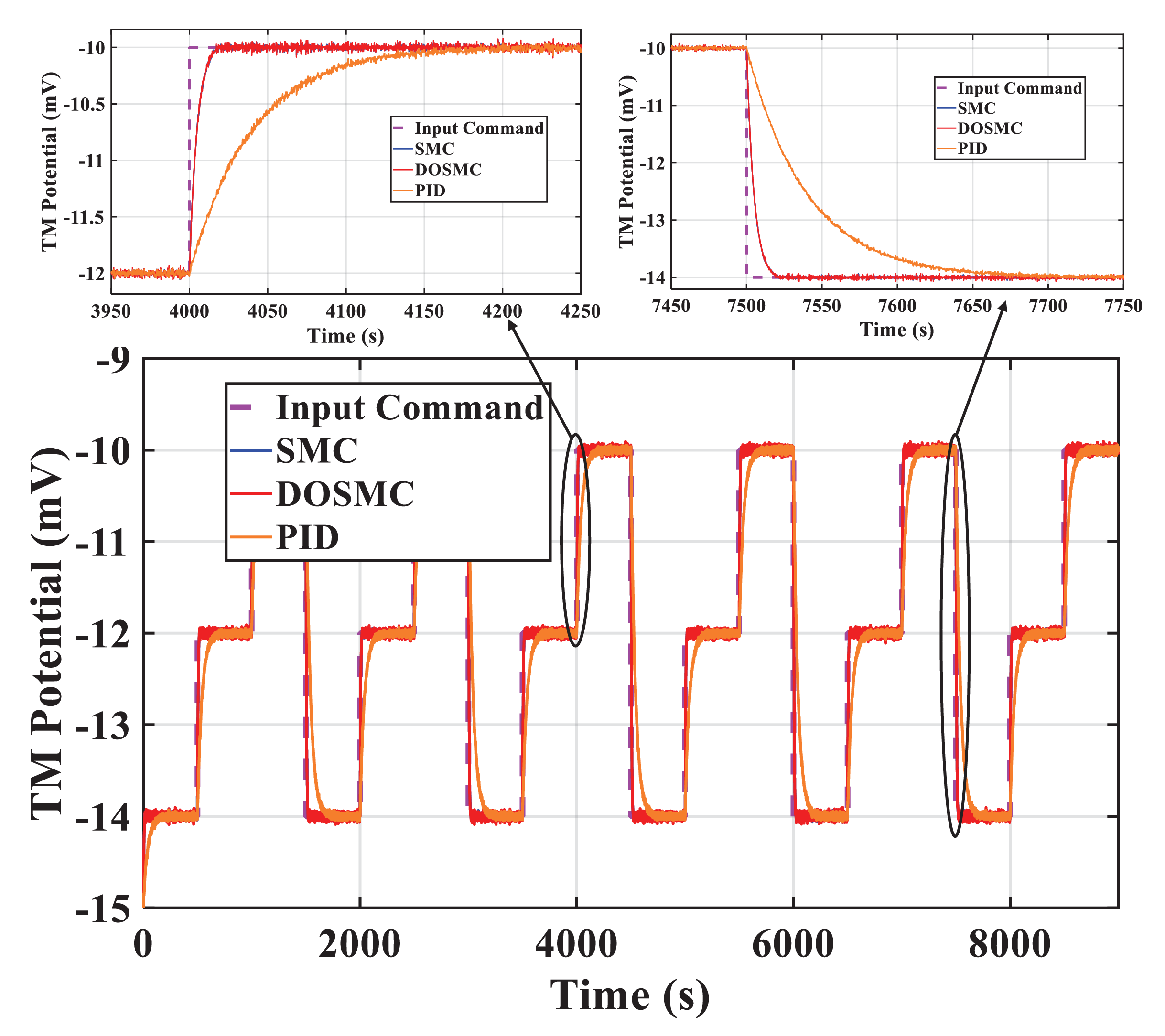}
}
\subfigure[The tracking errors of SMC, DOSMC and PID]{
\includegraphics[width=250pt]{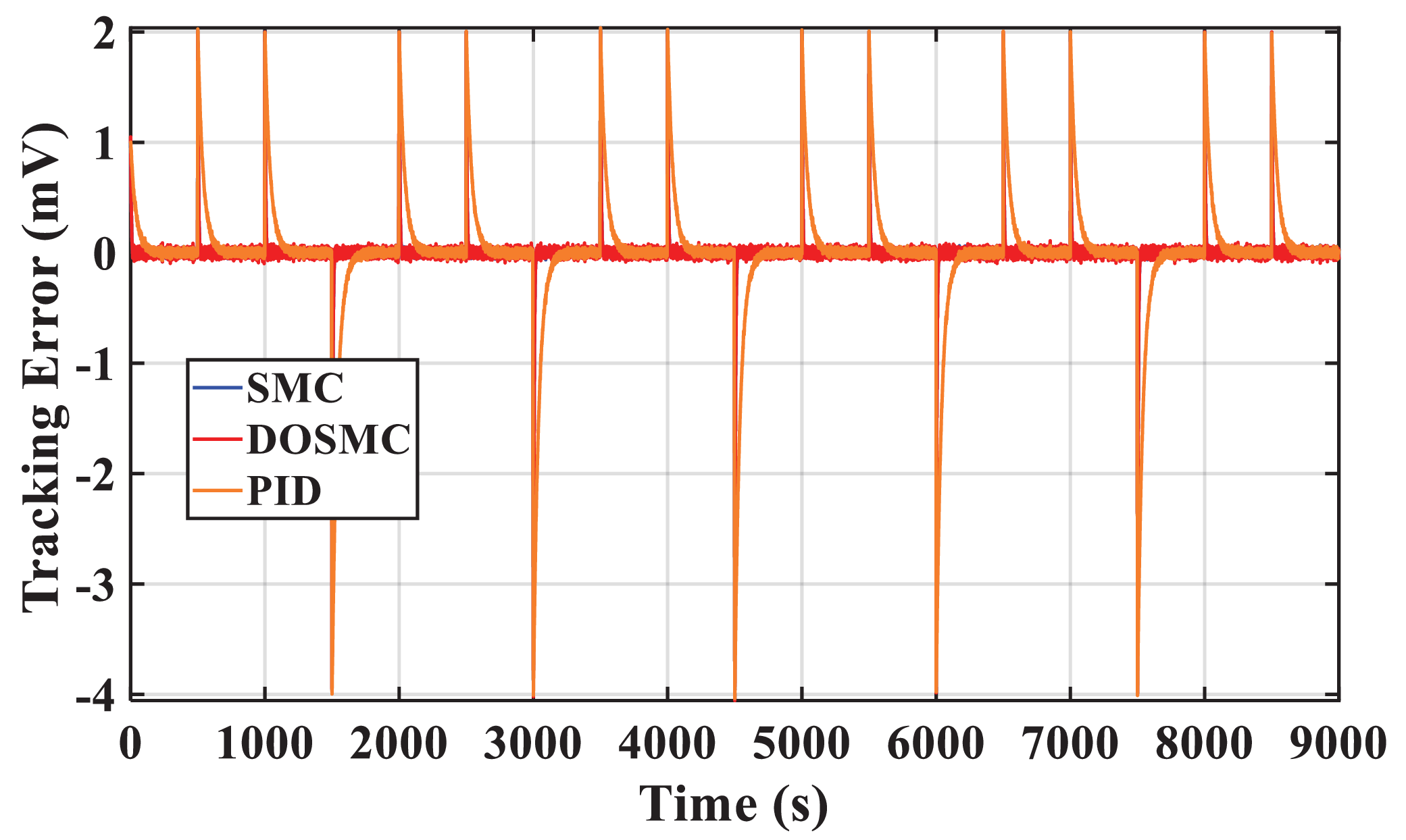}
}
\caption{\label{fig6} The closed-loop system response and tracking errors for the SMC, DOSMC and PID while tracking a series of step-input commands of different magnitudes under the normal condition. The amplitudes of step-input commands gradually increase from -14 mV to -10 mV in a step of 2 mV. The purple dashed line indicates the step-input commands, the blue, red and orange solid lines indicate the potential output of SMC, DOSMC and PID respectively. }
\end{figure}

Figure~\ref{fig6} shows the closed-loop system response(~\ref{fig6}(a)) and tracking errors(~\ref{fig6}(b)) for the SMC, DOSMC and PID  while tracking a series of step-input commands of different magnitudes under the normal condition. This total simulation includes six cycles, and the amplitude of the step-input commands in each cycle is gradually increased from -14 mV to -10 mV in a step (each step lasts 500 s) of 2 mV. These values are chosen because the acceleration distribution noise generated by the residual electric charge is below 2 $\times$ 10$^{-16}$ ${\rm{m/}}{{\rm{s}}^{\rm{2}}}{\rm{/H}}{{\rm{z}}^{{\rm{1/2}}}}$\cite{TIANQIN}, which satisfies the charge requirement for numerous space applications. The results reveal that all controllers are able to provide adequate closed-loop tracking performance. The control performance of SMC and DOSMC are almost the same because there is no disturbance term. The PID has a good tracking performance owing to the well-tuned control parameters for the parameter-fixed CMS. In addition, all controllers have a stable tracking error of the same order of magnitude (<0.1 mV).

\subsection{\label{sec:level8} Parameters variation}
As mentioned before, the variations in the parameters such as external charging rate, UV power and quantum yield affects the CMS performance, thus it is essential to evaluate specific impact of parameters variation on the stability and robustness of the charge control.

\begin{figure}
 \subfigure[Charging rate caused by a space radiation environment is set as a vector signal of the sequence stair, $\vec \alpha $= (20, 200, 3200, 16000, 40000, 80000) ]{
\includegraphics[width=250pt]{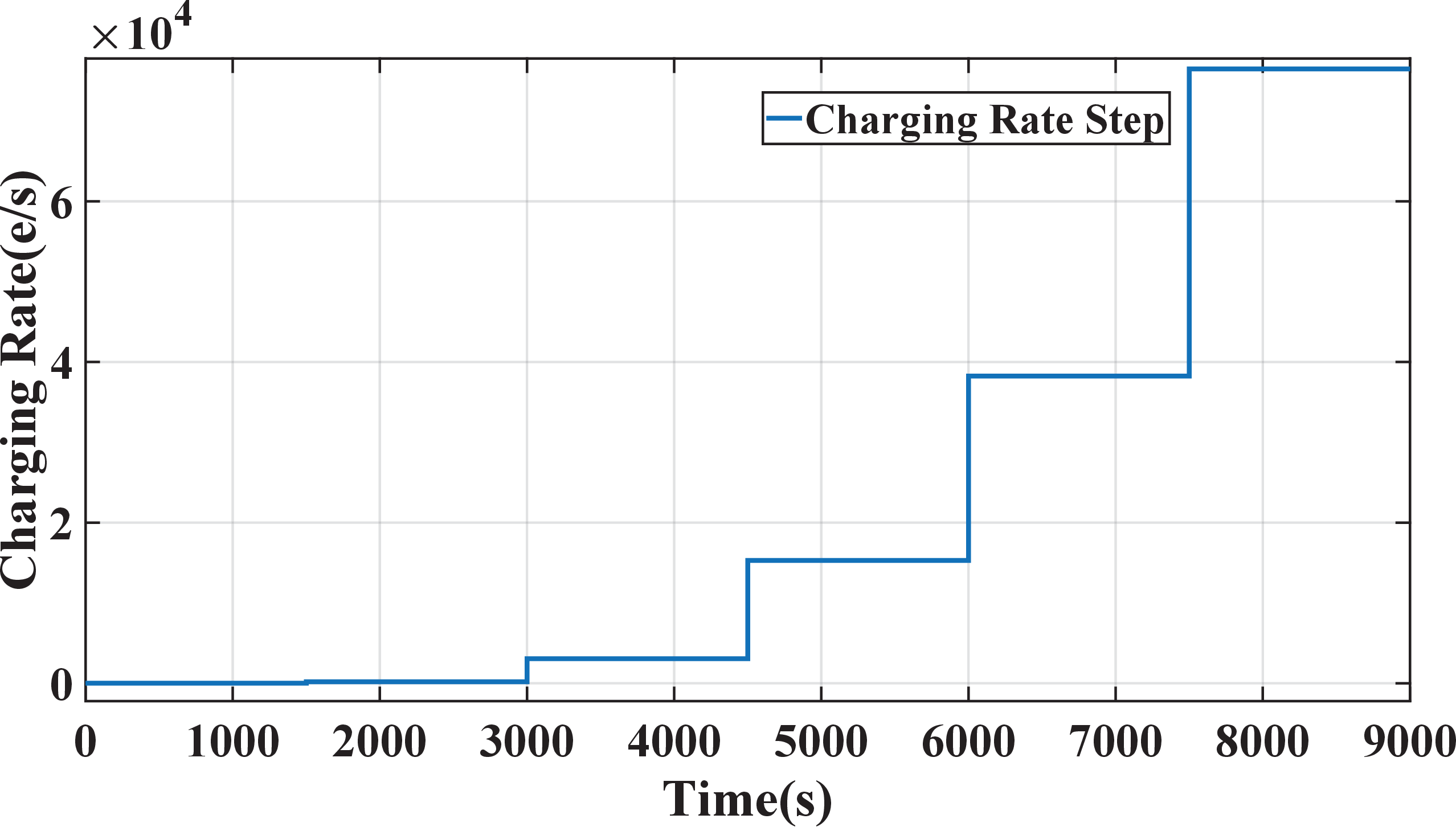}
}
\subfigure[Modified experimental data of the UV power attenuation of device SET-240-03 extracted from \cite{Hollington_2017}. The blue cross points indicate the experimental time series data, and the red solid line is the exponential fitting curve.]{
\includegraphics[width=250pt]{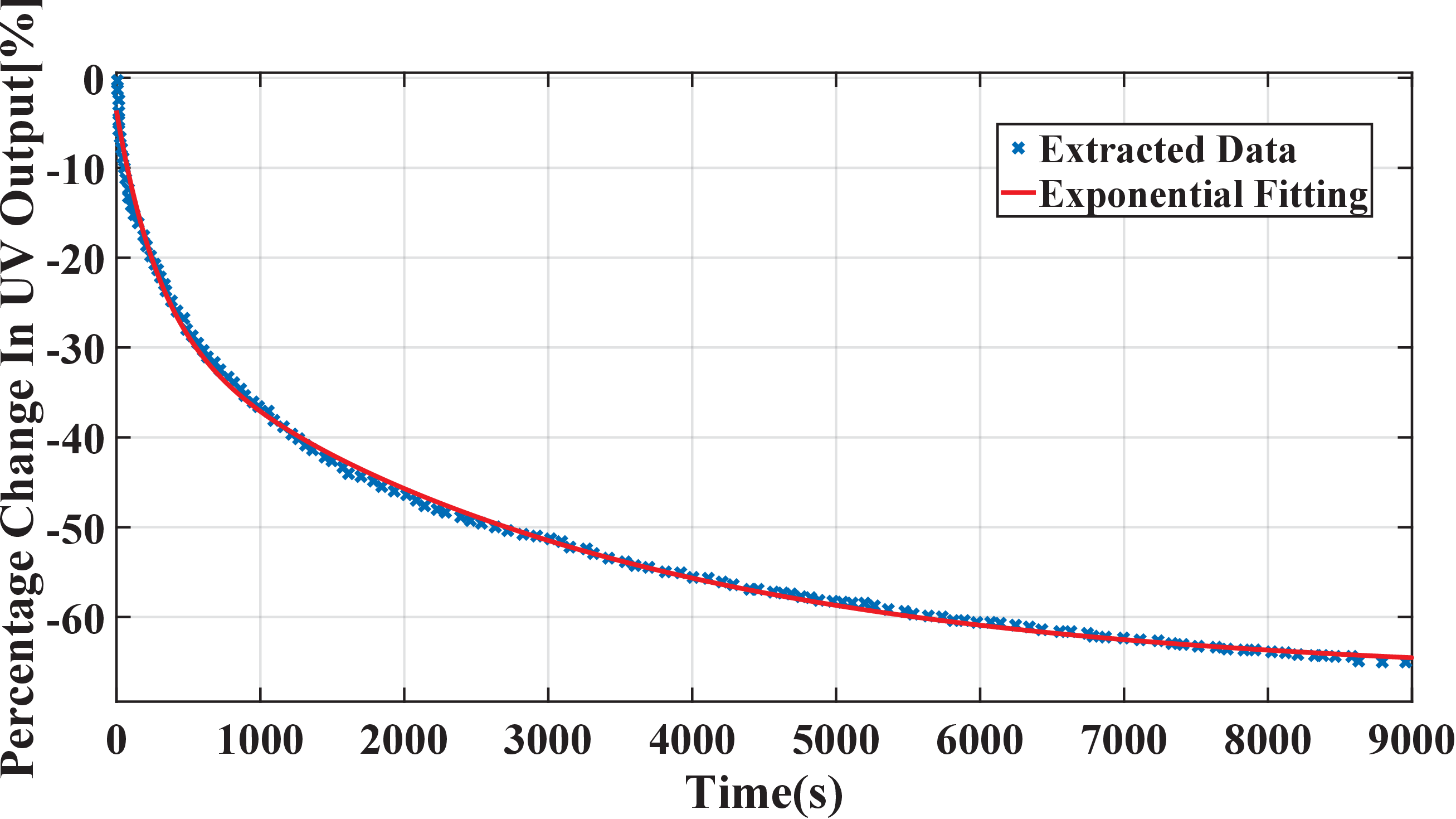}
}
\subfigure[Modified experimental data of the quantum yield variation of surface MTK-010 extracted from \cite{Wass_2019}. The blue cross points indicate the experimental time series data, and the red solid line is the exponential fitting model.]{
\includegraphics[width=250pt]{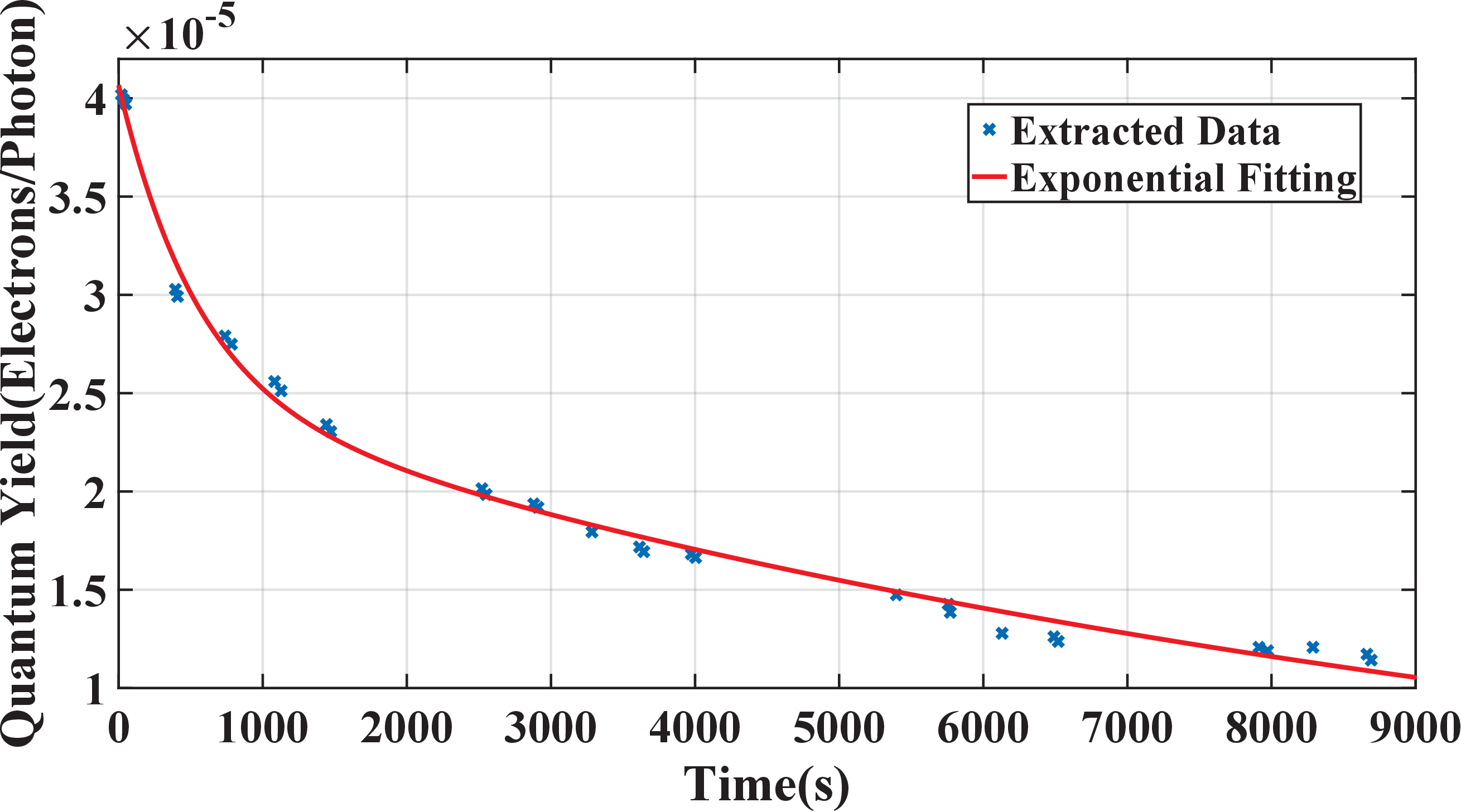}
}
\caption{\label{fig7} parameters variation data. (a) external charging rate variation. (b) UV power attenuation. (c) Quantum yield variation }
\end{figure}

First, the TM charging rate generated by GCR is nearly 25 +e/s for solar maximum conditions, increasing to 50 +e/s for solar minimum, and the charging rates expected from the two SEP events at a peak flux are estimated to be 87 +e/s and 68000 +e/s, respectively. Based on the above data, the charging rate caused by the space radiation environment was set as a vector signal of the sequence stair, $\vec \alpha $= [20, 200, 3200, 16000, 40000, 80000], and each step lasted 1500 s so that the total simulation time was 9000 s, as shown in Figure~\ref{fig7}(a).

Second, since a majority of space applications have a multiyear lifetime, so the aging problem of UV LED devices must be considered. In \cite{Hollington_2017}, experimental results of the UV power attenuation were obtained. Tests were performed at realistic output levels for both fast and continuous discharging in either a DC or pulsed mode of operation. The results showed that the output power of the UV light generated by UV LEDs would decrease by as much as 66 $\%$ for SET-240 devices and 10 $\%$ for CIS-250 devices after each device was turned on and off for about 50 days. Therefore, the effect of gradually reducing the UV output power on the performance of closed-loop charge control must be evaluated. The experimental data of device SET-240-03 in \cite{Hollington_2017} were  extracted, and the experimental time was scaled down to 9000 s to reduce the simulation time and make the simulation results more obvious. The modified experimental data and the exponential fitting curve used to describe the aging process of UV LED devices in the simulation are shown in Figure~\ref{fig7}(b).

Finally, surface properties, such as quantum yield, vary on different surfaces, or even at different times on the same surface, owing to changes in temperature, pressure, and air contamination. Therefore, the variation in the quantum yield should be considered. In \cite{Wass_2019}, a study of the photoelectric emission properties of 4.6 $\times$ 4.6 cm$^{2}$ gold-plated surfaces was reported. This surface is representative and is used in typical satellite applications with a film thickness of 800 nm and different surface roughnesses. The experimental data of surface MTK-010 (roughness: about 10 nm) were extracted, and the experimental time was also scaled down to 9000 s. Figure~\ref{fig7}(c) shows the modified experimental data and the exponential fitting curve of the saturated quantum yield data at normal incidence on surface MTK-010.

\begin{figure}
 \subfigure[The closed-loop system response for SMC, DOSMC and PID. Top-left: magnified view of the rising edge of the control performance. Top-right: magnified view of the falling edge of the control performance.]{
\includegraphics[width=250pt]{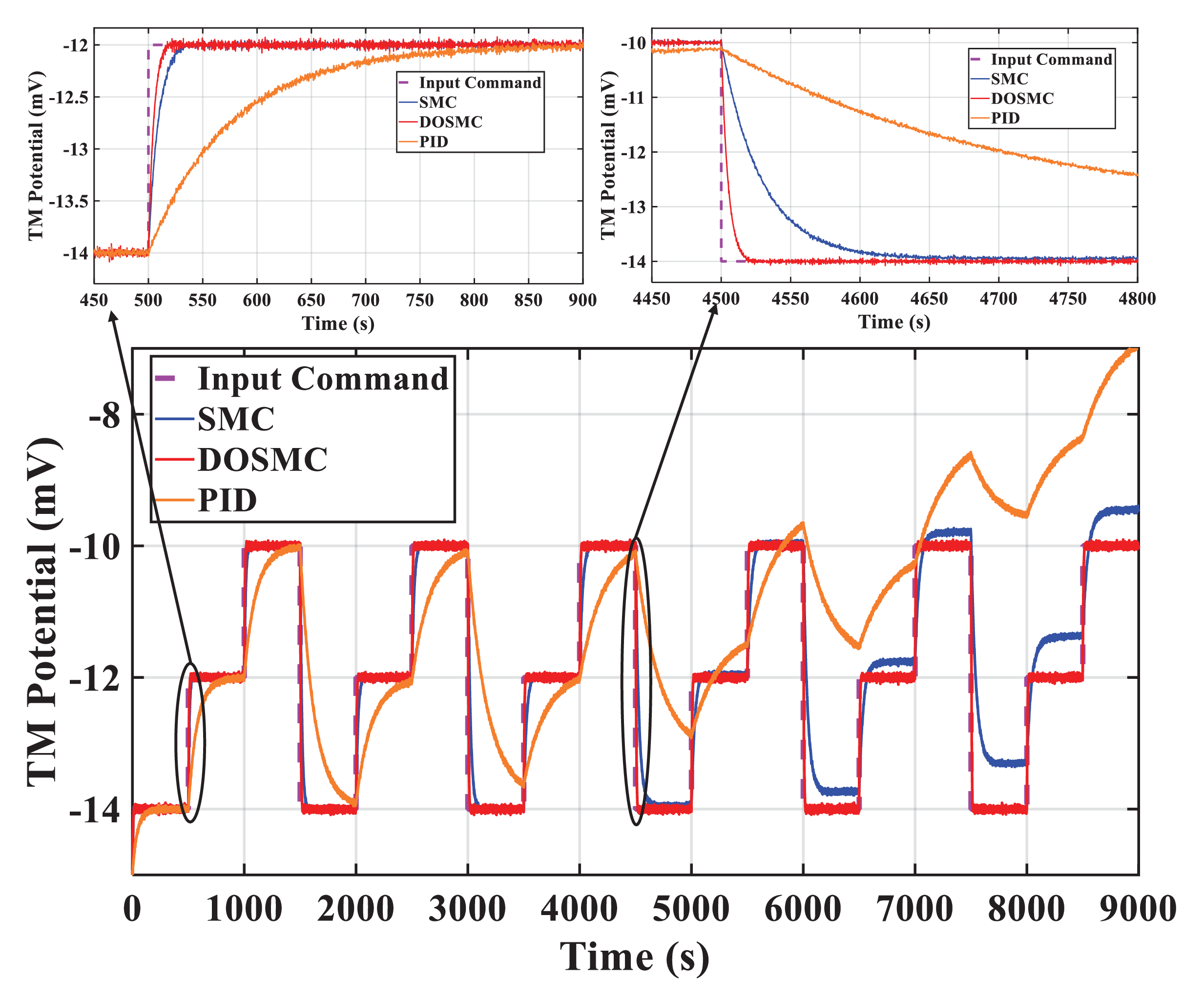}
}
\subfigure[The tracking errors of SMC, DOSMC and PID]{
\includegraphics[width=250pt]{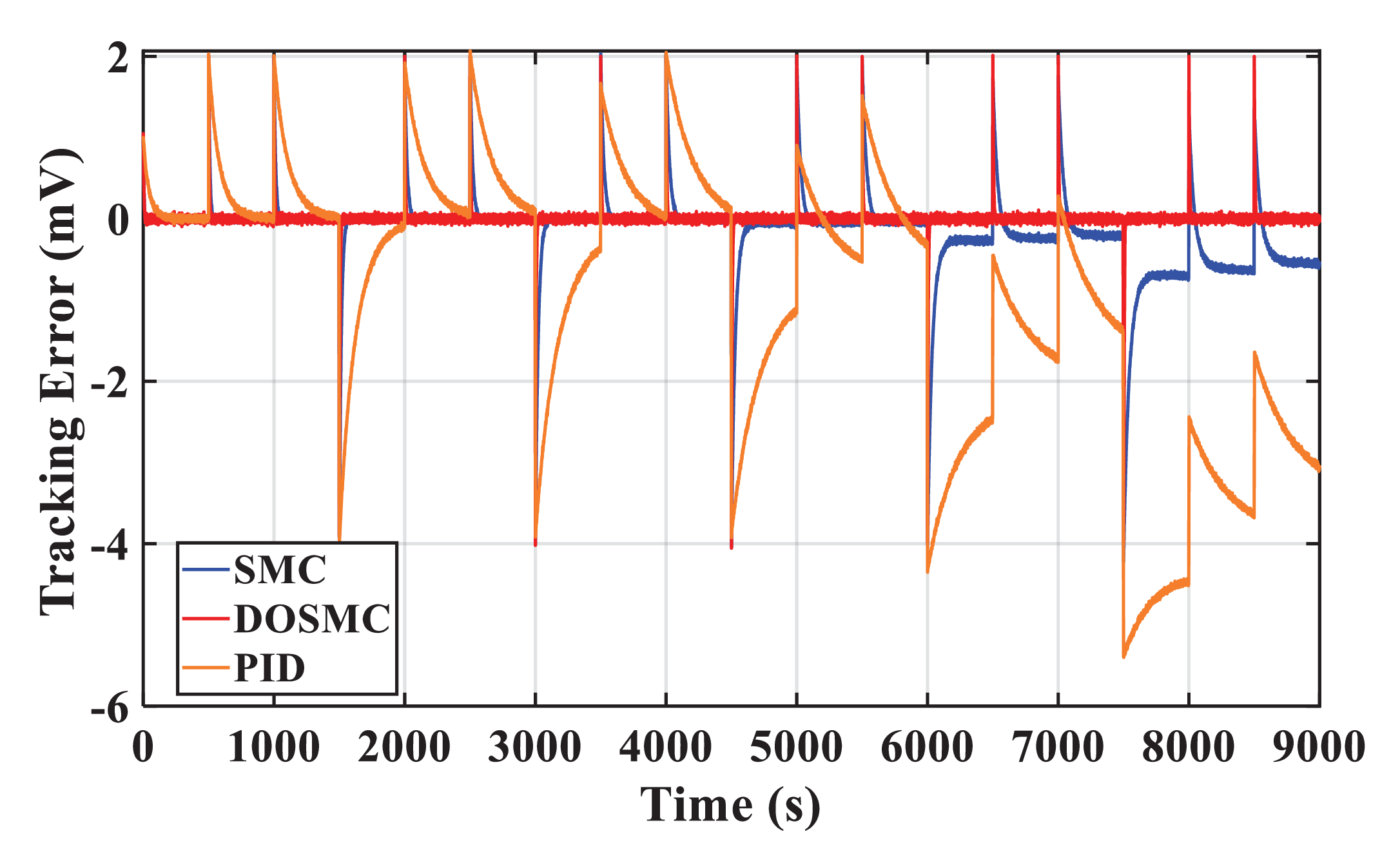}
}
\caption{\label{fig8} The closed-loop system response and tracking errors for the SMC, DOSMC and PID while tracking a series of step-input commands of different magnitudes under the condition of parameters variation. The amplitudes of step-input commands gradually increase from -14 mV to -10 mV in a step of 2 mV. The purple dashed line indicates the step-input commands, the blue, red and orange solid lines indicate the potential output of SMC, DOSMC and PID respectively.}
\end{figure}

The simulation responses of the overall closed-loop system under this abnormal condition are demonstrated in Figure~\ref{fig8}(a).  For the DOSMC, the response of the system still follows the expected potential well. For the SMC, the output TM potential can basically track the input command during but significant tracking errors occur in the last two cycles. As for The PID, the output TM potential can barely track the input command only during the initial two cycles and then gradually deviates from the target potential afterward with the change in the parameters. Figure~\ref{fig8}(b) shows the tracking errors of the TM potential output for these three controllers. The final tracking error of the SMC and PID reaches about 0.5 mV and 3 mV respectively, whereas that of the MRAC is almost the same as that under the normal condition.

\subsection{\label{sec:level9} Unknown disturbance}
\begin{figure}
 \subfigure[The closed-loop system response for SMC, DOSMC and PID. Top-left: magnified view of the rising edge of the control performance. Top-right: magnified view of the falling edge of the control performance.]{
\includegraphics[width=250pt]{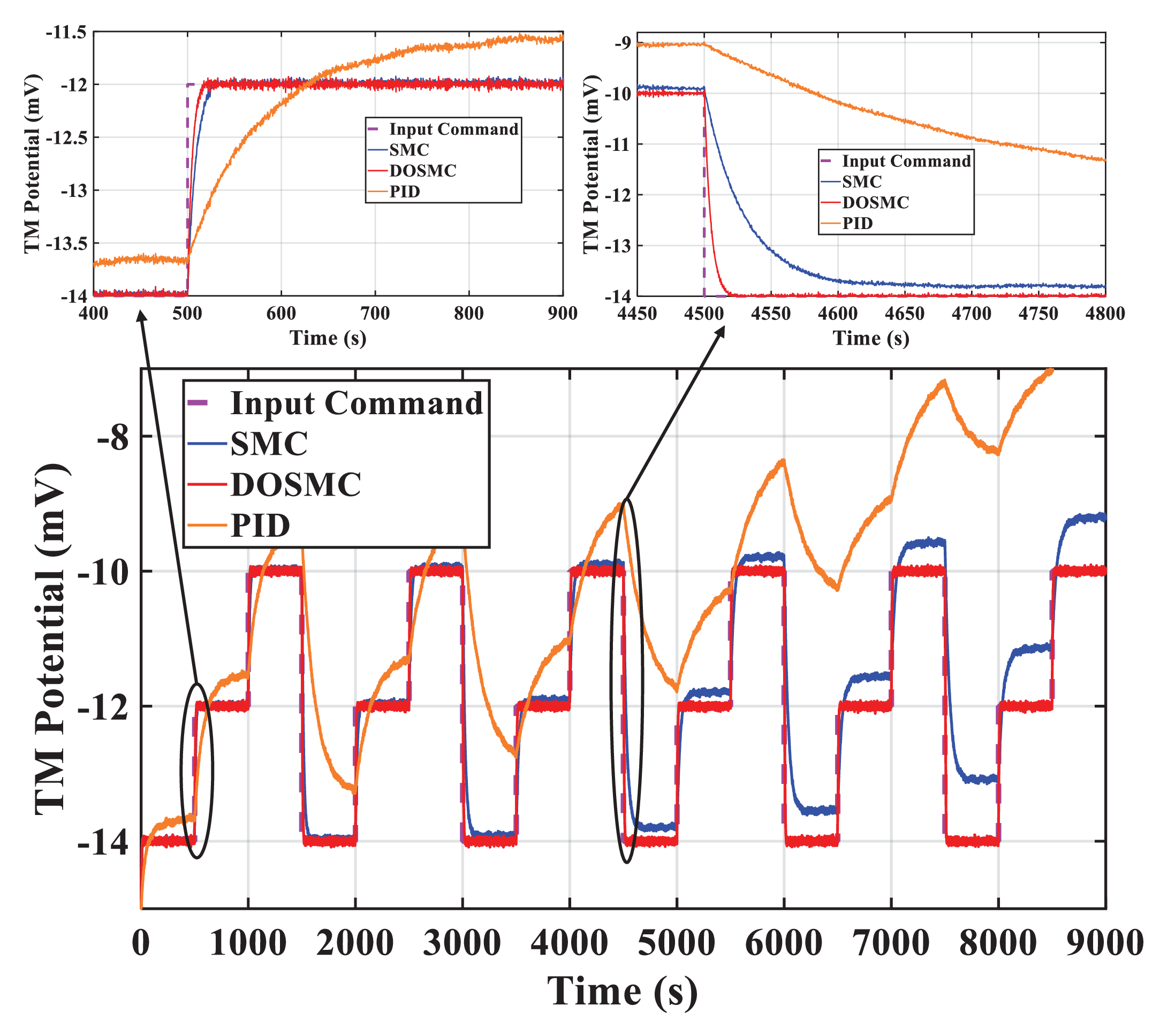}
}
\subfigure[The tracking errors of SMC, DOSMC and PID]{
\includegraphics[width=250pt]{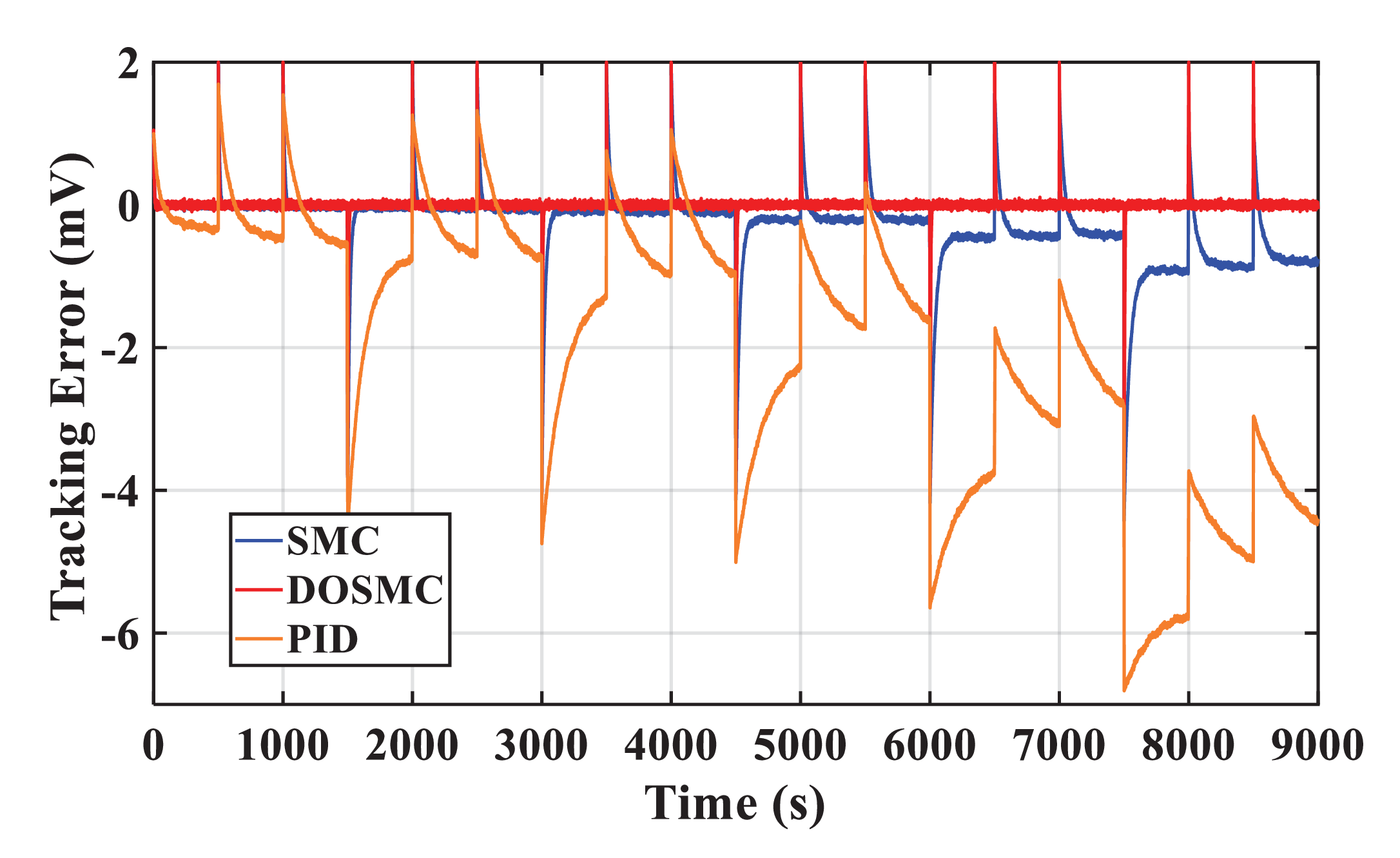}
}
\caption{\label{fig9} The closed-loop system response and tracking errors for the SMC, DOSMC and PID while tracking a series of step-input commands of different magnitudes under the condition of parameters variation and disturbance term. The amplitudes of step-input commands gradually increase from -14 mV to -10 mV in a step of 2 mV. The purple dashed line indicates the step-input commands, the blue, red and orange solid lines indicate the potential output of SMC, DOSMC and PID respectively.}
\end{figure}

To evaluate the stability and robustness of DOSMC under unknown and unmodeled disturbances, a disturbance term is introduced and set as a sinusoidal signal with a period of 100 s, an amplitude of 10$^{-2}$ mV/s and a bias of 5$\times$10$^{-2}$ mV/s, and all the other parameters
remained unchanged. The simulation responses of the overall closed-loop system for the SMC, DOSMC and PID under this abnormal condition are shown in Figure~\ref{fig9}(a). The DOSMC consistently maintains good tracking performance while the other two control methods cannot precisely control the TM potential. Figure~\ref{fig9}(b) presents the tracking errors of the TM potential output for all controllers. The tracking error of DOSMC remains consistent with the previous results, approximately at the order of 10$^{-4}$. On the contrary, the tracking errors of SMC and PID increase to approximately 0.8 mV and -4 mV, respectively, which cannot be ignored during the TM charge control process in high-precision space missions.

\section{Conclusion}
In this study, we introduced a DOSMC approach tailored for the charge management system (CMS) of contact-free UV discharge via photoemission, designed for ultrasensitive space-based gravitational missions. The findings demonstrate that this method effectively mitigates the unpredictability and variability of physical parameters and unpredictable interferences, thereby enhancing closed-loop tracking performance and robustness in charge control. First, a physics-based model of the discharge process with high charging/discharging rate that considers the electron exchange between a TM and an opposing EH in an inertial sensor was proposed. It successfully described the TM equilibrium potentials in the system and their dependance on the UV illumination properties and applied DC voltages. After that, the theoretical derivations of both SMC and DOSMC are presented, demonstrating that under external unknown disturbances, DOSMC can effectively suppress chattering, thereby improving the system's stability and robustness compared to SMC. Finally, the influence of parameter variation and unknown disturbance  in the case of long-term experiments was considered and evaluated in the simulation. This included abrupt changes in the external charging rates generated by GCR or SEP, device degradations of a deep UV LED light source, which is a candidate technology for future gravitational missions, and quantum yield variation of gold-coated surfaces used in space inertial sensors. Furthermore, the above-mentioned operations were repeated for a well-tuned PID controller and SMC. The results indicated that the DOSMC  precision can reach 0.1 mV under uncertainties and unpredictable interferences, which is superior to that of a classic PID controller and SMC.

This work aims to solve the unpredictability and variability of the external charging rate, UV output power, and surface quantum yields as well as unknown disturbance, which are key uncertainties in understanding the discharge properties of CMSs. The analysis of simulations described here provides a robust method for controlling the TM potential of CMSs with parametric uncertainties and Unmodeled perturbations, and is applicable to the development of CMSs for numerous space gravitational missions.


\begin{acknowledgments}
This work was supported by the National Natural Science Foundation of China (Grant number 12105250).
\end{acknowledgments}

\bibliography{ref}

\end{document}